\documentclass[pr, twocolumn, superscriptaddress]{revtex4}  
\usepackage{amssymb}
\usepackage{amsmath}
\usepackage{bbm}
\usepackage{graphicx}
\usepackage{color}
 
\begin{document}

\author{Philipp Werner}
\affiliation{Department of Physics, University of Fribourg, 1700 Fribourg, Switzerland}
\author{Aaram J. Kim}
\affiliation{Department of Physics and Chemistry, DGIST, Daegu 42988, Korea}
\author{Lei Geng}
\affiliation{Department of Physics, University of Fribourg, 1700 Fribourg, Switzerland}

\title{$\eta$-pairing in metallic and particle-hole asymmetric systems}

\date{\today}

\hyphenation{}

\begin{abstract}
Light-induced superconducting-like states have been reported in several classes of correlated materials. From a theoretical point of view, the induction of $\eta$-pairing is a promising route to nonthermal superconductivity. Numerical studies of photo-doped Mott systems revealed $\eta$-pairing states with very high effective critical temperatures. These investigations were however restricted to particle-hole symmetric states in large-gap Mott insulators, while the experiments were performed on strongly correlated metallic systems. It is thus relevant to explore if  $\eta$-pairing also exists in non-particle-hole symmetric setups and in photo-excited metallic states. Here we use steady-state nonequilibrium dynamical mean field theory combined with a strong-coupling impurity solver up to third order to investigate this issue. We find that in the strongly correlated regime with large Mott gap, and for low effective doublon and holon temperatures, $\eta$-pairing is robust against changes in the total filling and an imbalance in the doublon and holon density. An asymmetry in the effective doublon and holon temperatures can however strongly suppress the order parameter. In photo-doped metallic systems with a three-peak structure in the local density of states, $\eta$-pairing can be realized in set-ups with positive doublon and holon temperatures and a population inversion in the low-energy quasi-particle band. 
\end{abstract}

\maketitle


\section{Introduction}

Nonequilibrium control of material properties can give rise to novel devices like ultrafast and low-power memories \cite{Vaskivskyi2016,Mihailovic2021} or intriguing states of matter including metastable high-temperature superconductors \cite{Cavalleri2018}. Pump-probe experiments on cuprates \cite{Fausti2011,Kaiser2014} and organic compounds \cite{Mitrano2016,Buzzi2020,Budden2021} reported light-induced changes in optical properties which look consistent with a transient switching to a superconducting state, even at temperatures significantly above the equilibrium critical temperature $T_c$ of the material. In the case of cuprates, also an ultra-fast Meissner effect was measured using magneto-optical sampling \cite{Fava2024}. While the mechanisms behind these observations remain unclear, and questions about some aspects of the data analysis have been raised \cite{Dodge2023}, there have been several theoretical works discussing superconductivity induced by nonequilibrium driving. The possible mechanisms range from nonlinear phonon effects and phonon resonances \cite{Babadi2017,Murakami2017} to spin heating \cite{Tindall2019} and entropy-cooling \cite{Nava2018,Werner2019a}. A particularly robust strategy for superconductivity in highly-excited nonequilibrium systems is the realization of an $\eta$-pairing state \cite{Yang1989} through the condensation of doublon and holon type charge carriers in photo-doped Mott insulators \cite{Rosch2008,Kaneko2019,Werner2019b,Li2020,Geng2026}. The existence of  $\eta$-pairing in an infinitely repulsive Hubbard model with only doublons and holons has been discussed in the cold atom context in Ref.~\onlinecite{Rosch2008}, while Kaneko and co-workers showed that such a phase can also be induced by applying appropriate laser pulses to a one-dimensional Hubbard model \cite{Kaneko2019}. Dynamical mean field theory (DMFT) \cite{Georges1996,Aoki2014} simulations and effective equilibrium descriptions \cite{Murakami2022} have subsequently been employed to map out the nonequilibrium phase diagram of photo-doped Mott systems and to demonstrate that $\eta$-pairing is robust over a wide photo-doping range \cite{Li2020,Geng2026,Murakami2022}, provided that the effective temperature $T_\text{eff}$ of the doublon and holon charge carriers is low enough. For systems with a bandwidth of about 2 eV and a comparable Mott gap, $T_{c,\text{eff}}$ was recently shown to reach values above 1000~K \cite{Geng2026}, making the realization of a transient $\eta$-pairing state in cuprates and other strongly-correlated compounds a not completely unrealistic scenario. The DMFT-based studies have also been extended to multi-orbital systems and spin-triplet $\eta$-pairing \cite{Ray2023,Ray2024}, and to nonthermal chiral superconducting states in geometrically frustrated systems \cite{Li2023}. 

While most existing studies of $\eta$-pairing have focused on particle-hole symmetric photo-doped Mott systems, the experimentally studied materials for which light-induced superconductivity has been reported are correlated metals which lack particle-hole symmetry. This raises the question if $\eta$-pairing can be potentially realized in such systems, at least as a short-lived transient state. The question is also conceptually important: the particle-hole symmetric Hubbard model has an exact $\eta$-spin SU(2) symmetry~\cite{Yang1989}, which provides the natural setting for discussing $\eta$ pairing.  But finite-momentum onsite pair condensation is not forbidden away from this high-symmetry limit. Particle-hole asymmetric systems therefore provide a setup to test the robustness of $\eta$ pairing under more general conditions.
Here, we address this topic using a steady-state formulation of nonequilibrium DMFT \cite{Li2021,Kim2025}. By imposing distribution functions with given effective doublon and holon temperatures \cite{Kuenzel2024,Geng2025} in particle-hole asymmetric gapped systems, as well as metallic states with Hubbard bands, we explore the robustness of the $\eta$-pairing phase beyond the particle-hole symmetric large-gap Mott insulator regime. 

The paper is organized as follows. In Sec.~\ref{sec_model} we describe the model and the DMFT self-consistency loop for the Bethe lattice. Section~\ref{sec_results} presents our results for particle-hole asymmetric setups, including chemically doped gapped systems and photo-doped Mott insulators with different effective temperatures of the doublons and holons. We also test if an $\eta$-pairing state can be realized as a nonequilibrium steady-state solution in the correlated metallic regime. The results are summarized in Sec.~\ref{sec_conclusions}. 

\begin{figure}[t]
\begin{center}
\includegraphics[angle=-90, width=\columnwidth]{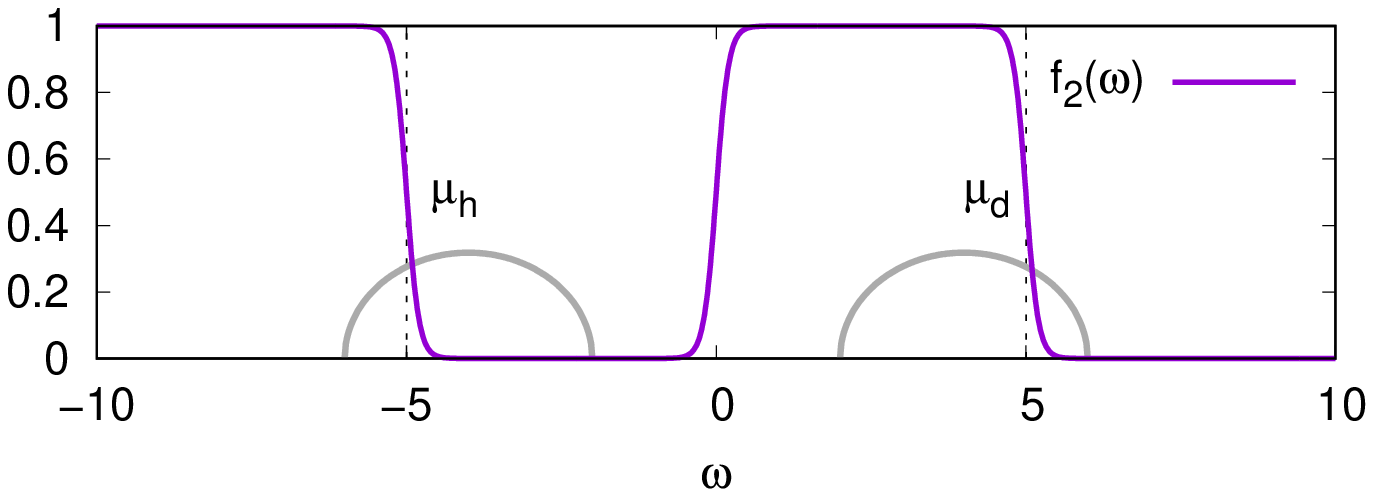}
\includegraphics[angle=-90, width=\columnwidth]{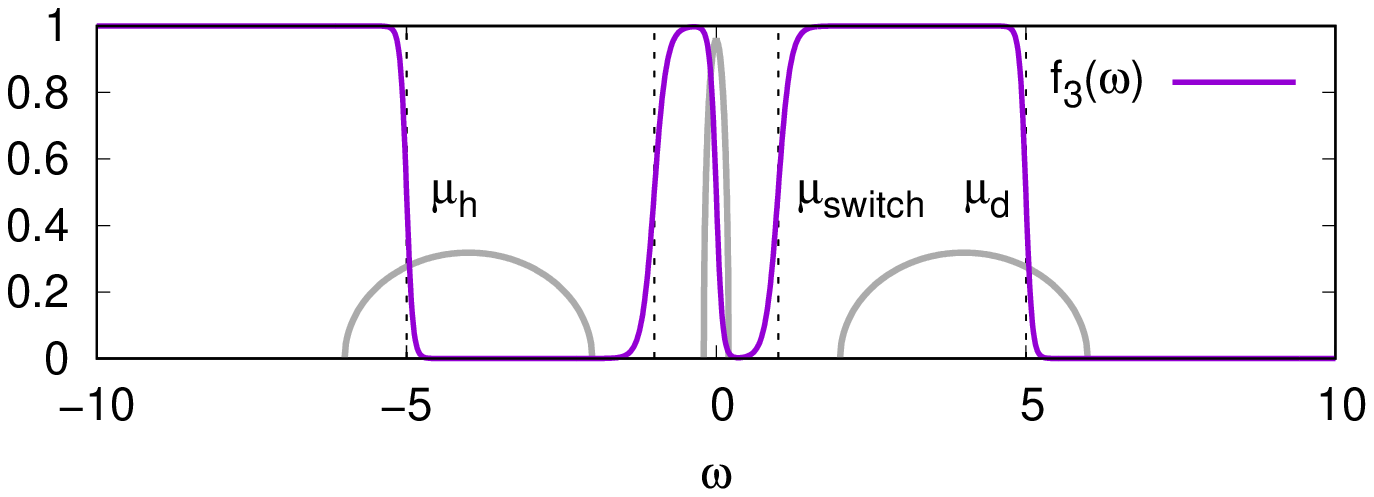}
\caption{Top panel: Illustration of a nonequilibrium distribution function $f_2(\omega)$ with two Fermi edges at $\mu_d$ and $\mu_h$ and a $1-f_\text{Fermi}$ type switching near $\omega=0$.
Bottom panel: Illustration of a nonequilibrium distribution function $f_3(\omega)$ with three Fermi edges at $\mu_d$, $0$ and $\mu_h$, and two $1-f_\text{Fermi}$ type switchings near $\omega=\pm\mu_\text{switch}$.
The gray lines are sketches of a Mott insulating (top panel) and metallic (bottom panel) density of states. 
}
\label{fig_f}
\end{center}
\end{figure}

\section{Model}
\label{sec_model}

We calculate the nonequilibrium steady-state properties of the single-band Hubbard model with local interaction $U$ on the Bethe lattice, using nonequilibrium DMFT \cite{Li2020,Li2021} in combination with a first order (noncrossing approximation, NCA \cite{Keiter1971}), second-order (one-crossing approximation, OCA \cite{Pruschke1989}), and third-order approximation (TOA) \cite{Kim2025,Geng2025} strong-coupling impurity solver. The nonequilibrium distribution function $f(\omega)$ corresponding to given effective doublon and holon temperatures is imposed by hand and the corresponding self-consistent steady-state DMFT solution calculated by iteration \cite{Kuenzel2024}. 
We consider two types of nonequilibrium distributions, as illustrated in Fig.~\ref{fig_f}. For gapped systems with $U$ larger than the critical value $U_c$ for the Mott transition at half-filling, we consider functions $f_2(\omega)$ with two Fermi-like distributions for the doublons (holons) in the upper (lower) Hubbard band. The energy positions of these Fermi edges are determined by the parameters $\mu_d$ and $\mu_h$. A smooth $1-f_\text{Fermi}$ type switching function near $\omega=0$ (in the gapped region) connects the two Fermi distributions. Unless otherwise noted, we use the same effective temperature for the switching function as for the doublons and holons. For metallic systems with $U<U_c$, we will additionally consider nonequilibrium distributions $f_3(\omega)$ with three Fermi edges at $\omega=\mu_d$, $0$ and $\mu_h$. In this case, the three Fermi distributions are connected by two switching functions at $\omega=\pm\mu_\text{switch}$.

In order to treat the $\eta$-pairing state, we use the Nambu formalism \cite{Georges1996} and define the Green's function
\begin{equation}
G(t,t')=\left(
\begin{tabular}{cc}
$G^{cc^\dagger}_{\uparrow\uparrow}(t,t')$ &  $G^{cc}_{\uparrow\downarrow}(t,t')$\\
$G^{c^\dagger c^\dagger}_{\downarrow\uparrow}(t,t')$ &  $G^{c^\dagger c}_{\downarrow\downarrow}(t,t')$
\end{tabular}
\right).
\label{eq_nambu}
\end{equation}
In terms of this Nambu Green's function, the self-consistency condition for the Bethe lattice reads \cite{Werner2019b}
\begin{equation}
\Delta(t,t')=v^2 G(t,t'),\label{eq_self}
\end{equation}
with $\Delta(t,t')$ the hybridization function of the effective impurity model \cite{Aoki2014} and $v$ the renormalized hopping parameter of the infinitely connected lattice \cite{Metzner1989}, which sets the bandwidth $W=4v$ of the noninteracting model. (Because of the staggered nature of the $\eta$-pairing order parameter, there  are no $\sigma_z$ factors in the self-consistency equation \cite{Werner2019b}.) 
The $\eta$-pairing order parameter corresponds to the equal-time value of the off-diagonal components. In practice, it is obtained from the pseudo-particle density matrix $\rho_{\rm pp}$ as $\mathrm{Tr}\left[\rho_{\rm pp}\, c_\uparrow c_\downarrow\right]$, where $c_\sigma$ is the annihilation operator for spin $\sigma$. 

In this study, we use $v$ as the unit of energy. $A^<(\omega)=\frac{1}{2\pi}\text{Im}G_{\uparrow\uparrow}^{cc^\dagger,<}(\omega)$ denotes the occupation function and $A(\omega)=-\frac{1}{\pi}\text{Im}G_{\uparrow\uparrow}^{cc^\dagger,\text{ret}}(\omega)$ the spectral function of the $\uparrow\uparrow$ component in the time-translation invariant steady state. The (nonequilibrium) distribution function is defined as $f(\omega)=A^<(\omega)/A(\omega)$.

\begin{figure}[t]
\begin{center}
\includegraphics[angle=-90, width=\columnwidth]{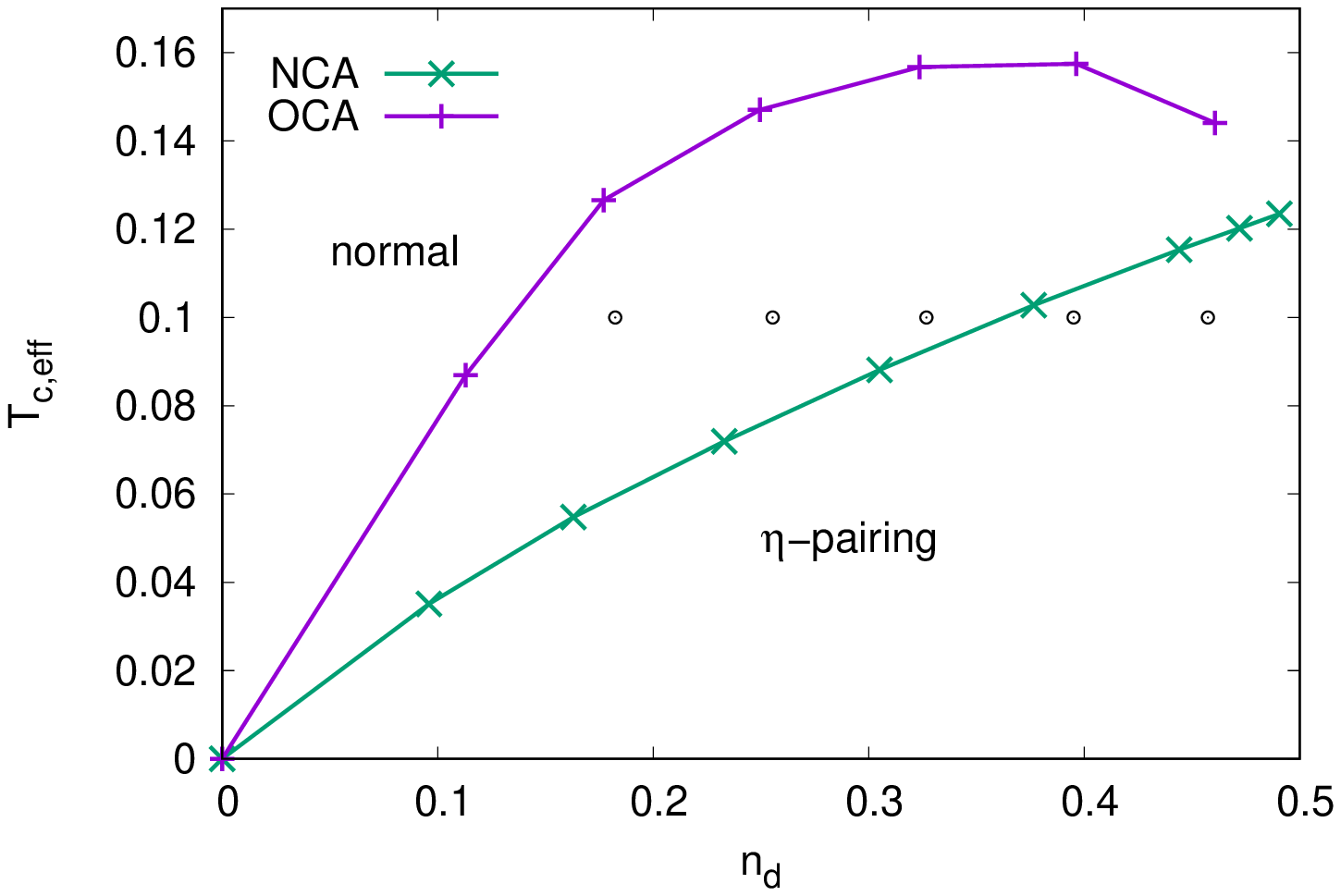}
\caption{Phase diagram of the half-filled Hubbard model with $U=8$ in the space of doublon density (= holon density) and effective temperature $T_\text{eff}$. 
The green and violet crosses show the effective critical temperatures $T_{c,\text{eff}}$ calculated with NCA and OCA, respectively, while the black dots mark representative parameter sets used below to study the effect of particle-hole asymmetry.}
\label{fig_phase}
\end{center}
\end{figure}

\begin{figure*}[t]
\begin{center}
\includegraphics[angle=-90, width=0.32\textwidth]{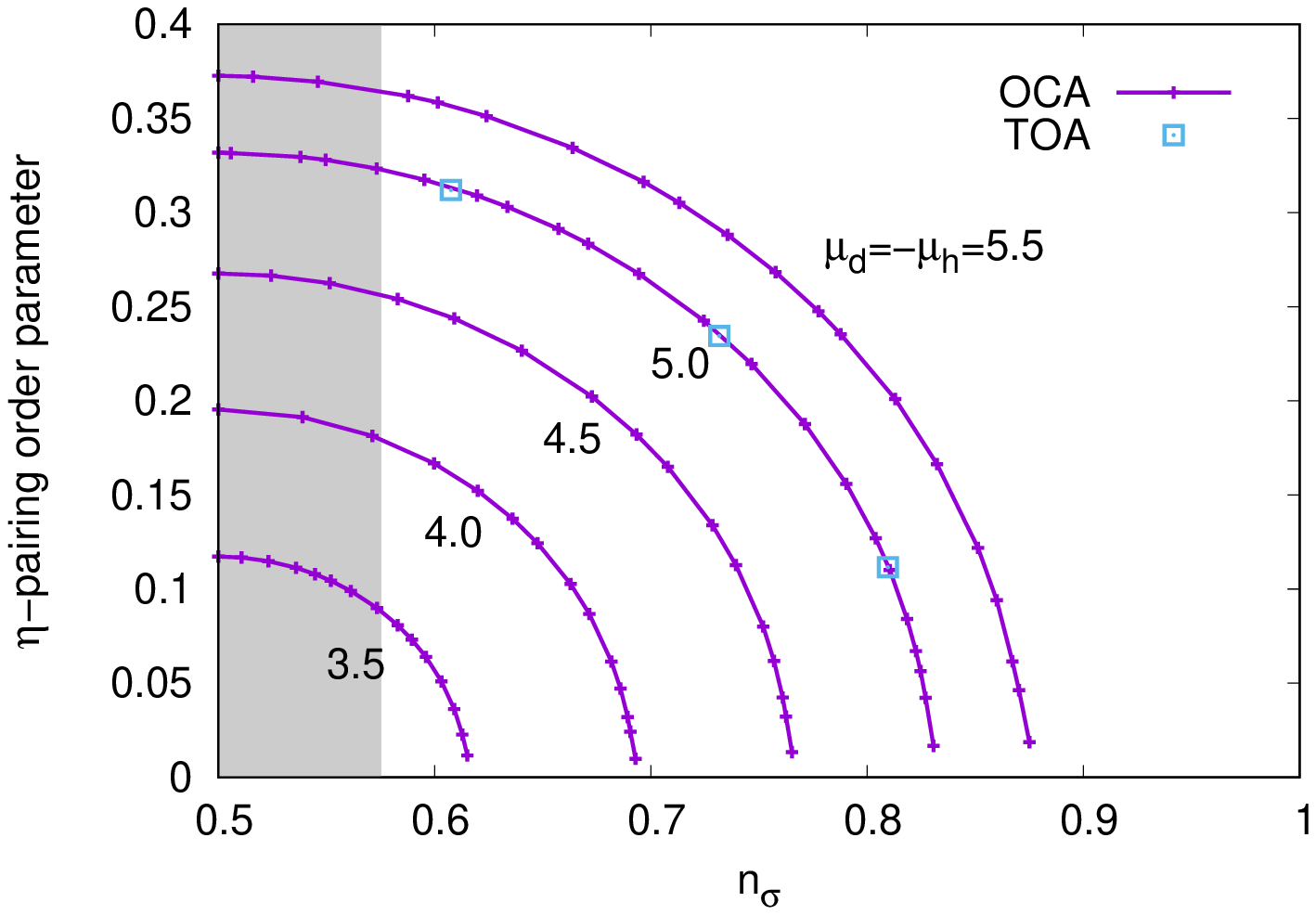}\hfill
\includegraphics[angle=-90, width=0.32\textwidth]{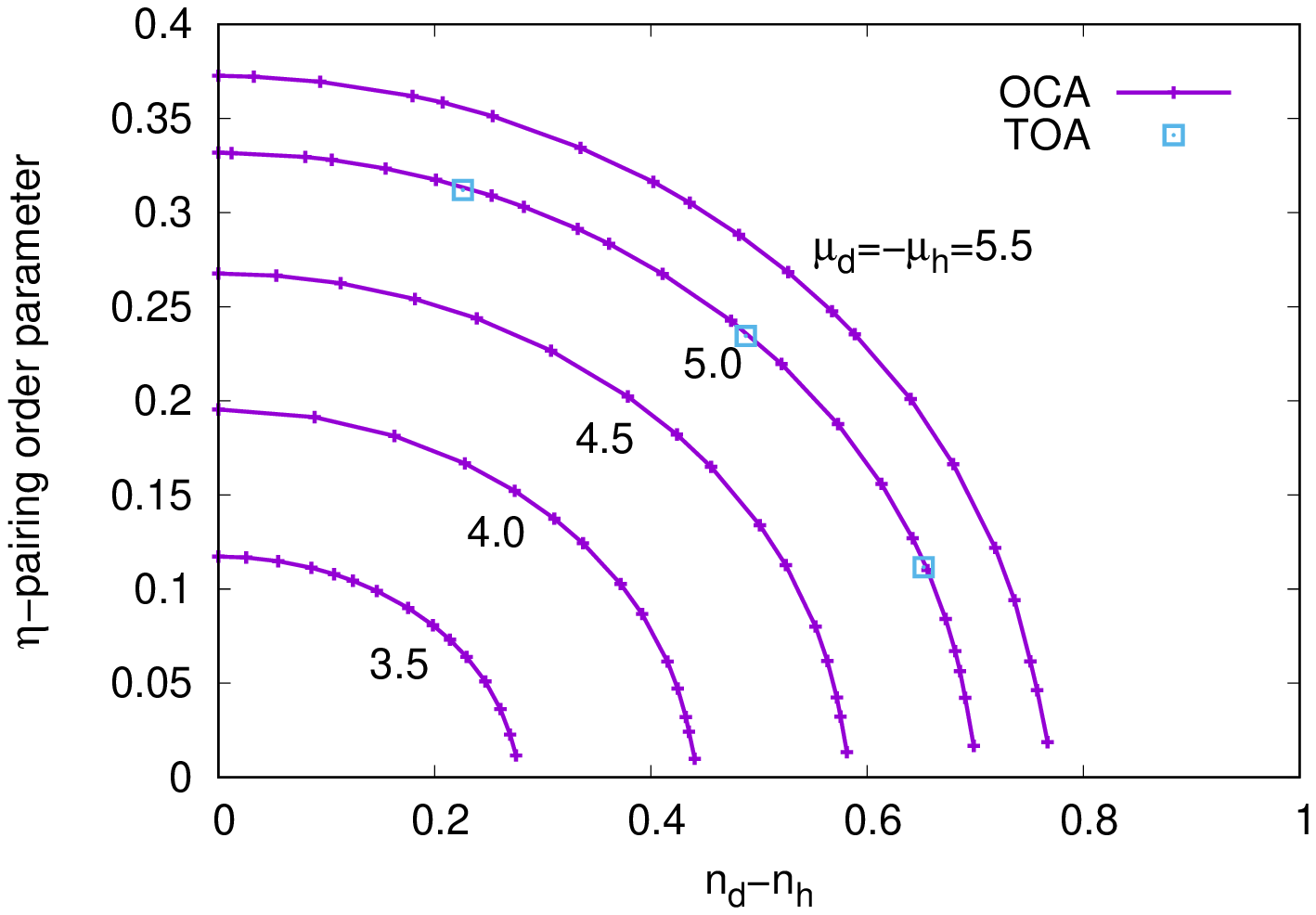}\hfill
\includegraphics[angle=-90, width=0.32\textwidth]{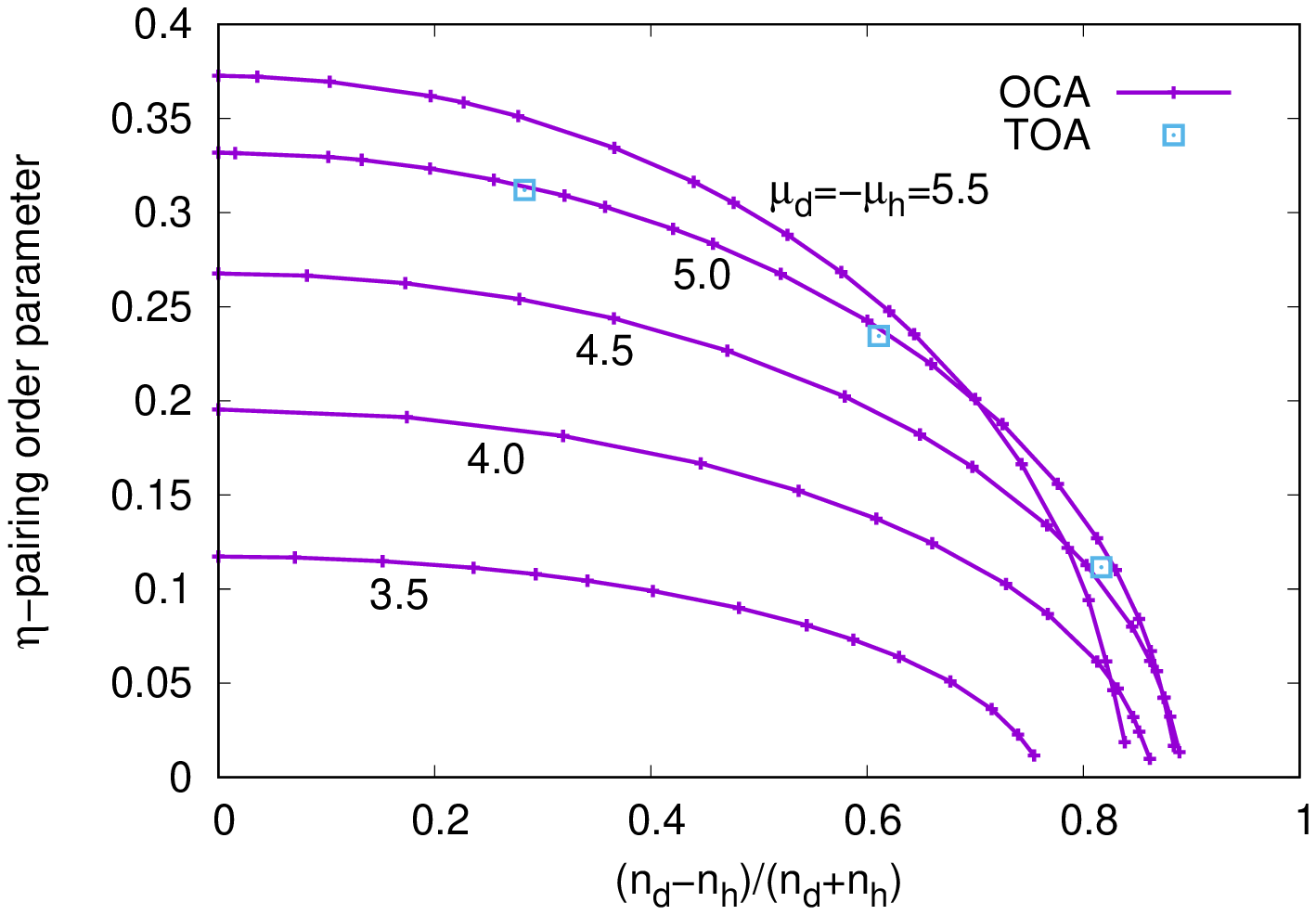}
\caption{$\eta$-pairing order parameter as a function of filling $n_\sigma$ (left panel), doublon-holon asymmetry $n_d-n_h$ (middle panel) and relative doublon-holon asymmetry $(n_d-n_h)/(n_d+n_h)$ (right panel) for the indicated values of $\mu_d=-\mu_h$ ($U=8$, $\beta_d=\beta_h=10$).  
The violet crosses are OCA results. For $\mu_d=-\mu_h=5.0$, we also show three data points obtained with the TOA.
}
\label{fig_op}
\end{center}
\end{figure*}

\section{Results}
\label{sec_results}

\subsection{Nonequilibrium phase diagram of the particle-hole symmetric system} 

First, we consider photo-doped Mott insulators with $U=8$. The $\eta$-pairing phase diagram for the particle-hole symmetric half-filled system is shown in Fig.~\ref{fig_phase}. The effective temperature $T_\text{eff}=1/\beta_\text{eff}$ determines the Fermi function like energy distribution of the doublons (holons) within the energy range of the upper (lower) Hubbard band, see $f_2(\omega)$ in Fig.~\ref{fig_f}. At the OCA level, the critical $T_{c,\text{eff}}$ for $\eta$-pairing is significantly enhanced compared to the NCA result. The phase boundary shows a dome-like shape with a maximum $T_{c,\text{eff}}$ near doublon (=holon) density $n_d=0.35$, and a roughly linear decrease to zero for small $n_d$. This result is consistent with the phase diagram for the 2D Hubbard model in Ref.~\onlinecite{Geng2026}, which additionally showed that the TOA phase boundary is essentially on top of the OCA result. 
 
Note that the density of long-lived doublon type charge carriers $n_d$ (``doublons") is not identical to the double occupation $\mathrm{Tr}\big[\rho_{\rm pp}\, c^\dagger_\uparrow c_\uparrow c^\dagger_\downarrow c_\downarrow\big]$, since the latter also contains contributions from quantum fluctuations. We evaluate the doublon density by integrating the occupation function (lesser spectrum) over the $\omega\ge 0$ half-interval, $n_d=\int_0^\infty d\omega A^<(\omega)$, and similarly the holon density $n_h$ is obtained by integrating the greater spectrum over $\omega\le 0$, $n_h=\int_{-\infty}^0 d\omega A^>(\omega)$.

To put the effective temperatures in Fig.~\ref{fig_phase} into a context, let us choose a bandwidth $W=2$~eV ($v=0.25$) representative of La$_2$CuO$_4$, which in single-site DMFT studies is typically simulated with a Hubbard interaction $U\approx 1.5 W$ \cite{Comanac2008}. For this bandwidth, $T_\text{eff}/v=0.16$ corresponds to 930~K. While photo-doping concentrations of $n_d=0.3$-$0.4$ are unrealistic in a solid-state setup, even at $n_d=0.05$, the estimated $T_{c,\text{eff}}$ reaches room temperature. If the photo-doped doublons and holons couple to a cold lattice, it is thus conceivable that the electronic temperatures required for $\eta$-pairing could be realized in photo-doping experiments.

\subsection{Nonequilibrium steady-states with doublon-holon imbalance}
\label{subsec_doped}

In realistic materials, the photo-doping process generically involves excitations between low-energy correlated states and higher energy (filled or empty) bands. In the absence of specific symmetries, this will result in a photo-doped state with different doublon and holon densities, $n_d\ne n_h$. It is thus relevant to explore the robustness of the $\eta$-pairing state against such a doublon-holon imbalance. Note that these doublon-holon asymmetric transient states will generically not be half-filled either. Hence, we may also regard them as the result of photo-excitation within the low-energy manifold of a chemically doped Mott system. 

We numerically explore the doublon-holon imbalanced systems by fixing the effective chemical potentials of the doublons and holons symmetrically, $\mu_d=-\mu_h$, while shifting the chemical potential $\mu$ (which controls the total filling) away from the particle-hole symmetric value $\mu=U/2$. By increasing $\mu$, we change both the total filling per spin, $n_\sigma$, and the doublon-holon imbalance $n_d-n_h$. Figure~\ref{fig_op} plots the $\eta$-pairing order parameter from OCA calculations for $U=8$ as a function of $n_\sigma$ (left panel), $n_d-n_h$ (middle panel) and $(n_d-n_h)/(n_d+n_h)$ (right panel). Here, we fix the effective inverse doublon and holon temperature to $\beta_d=\beta_h=10$ and choose $\mu_d=-\mu_h=3.5, 4.0, 4.5, 5.0, 5.5$ to control the photo-doping level, see the black circles in Fig.~\ref{fig_phase} for the particle-hole symmetric reference states. For $\mu_d=-\mu_h=5.0$ we also report three data points obtained with the computationally much more heavy TOA. The $\eta$-pairing order parameter is essentially converged at the OCA level. This was already shown for a particle-hole symmetric Hubbard model in Ref.~\cite{Geng2026}, and apparently also holds in the particle-hole asymmetric case. 

\begin{figure*}[t]
\begin{center}
\includegraphics[angle=-90, width=0.66\columnwidth]{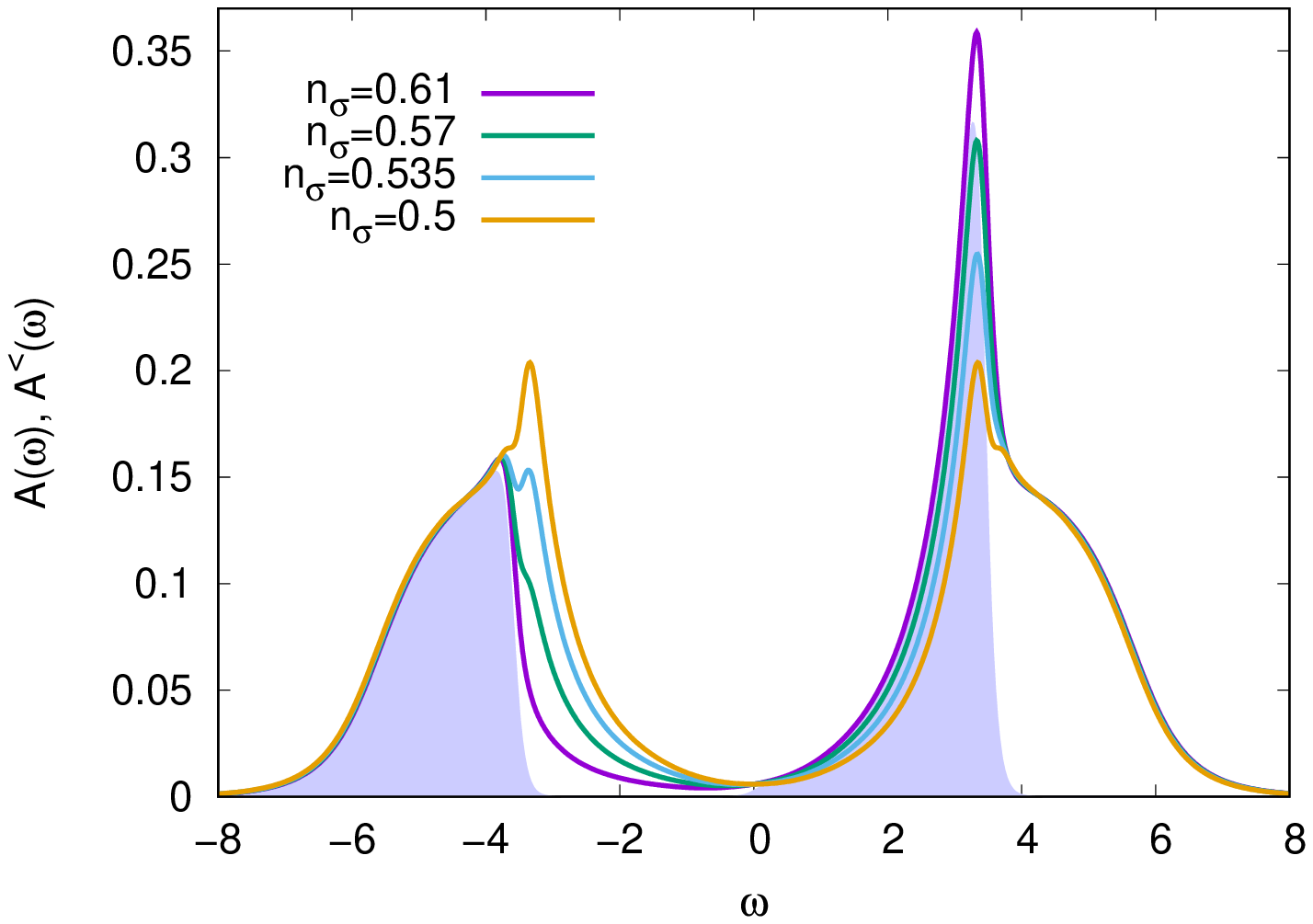}\hfill
\includegraphics[angle=-90, width=0.66\columnwidth]{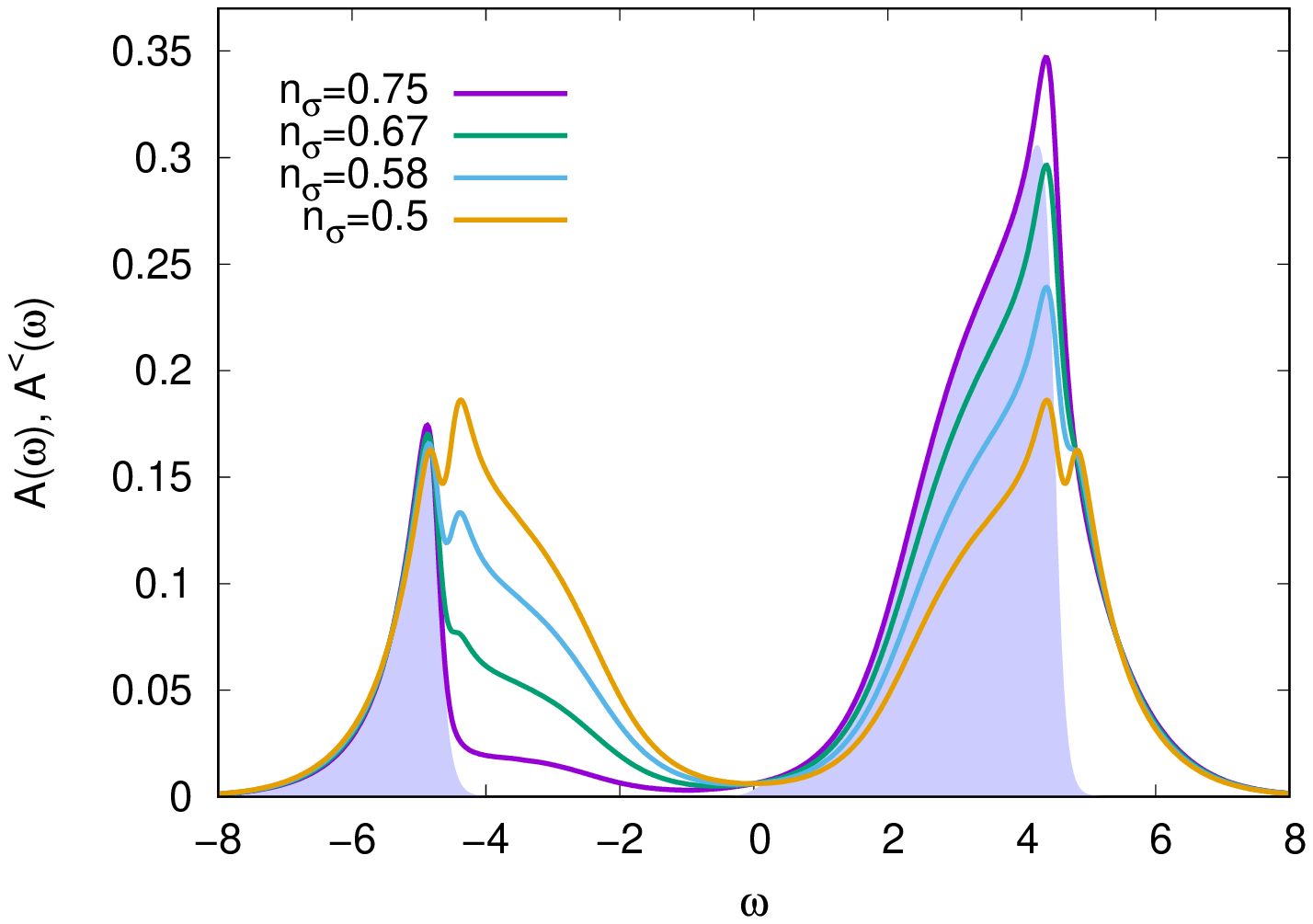}\hfill
\includegraphics[angle=-90, width=0.66\columnwidth]{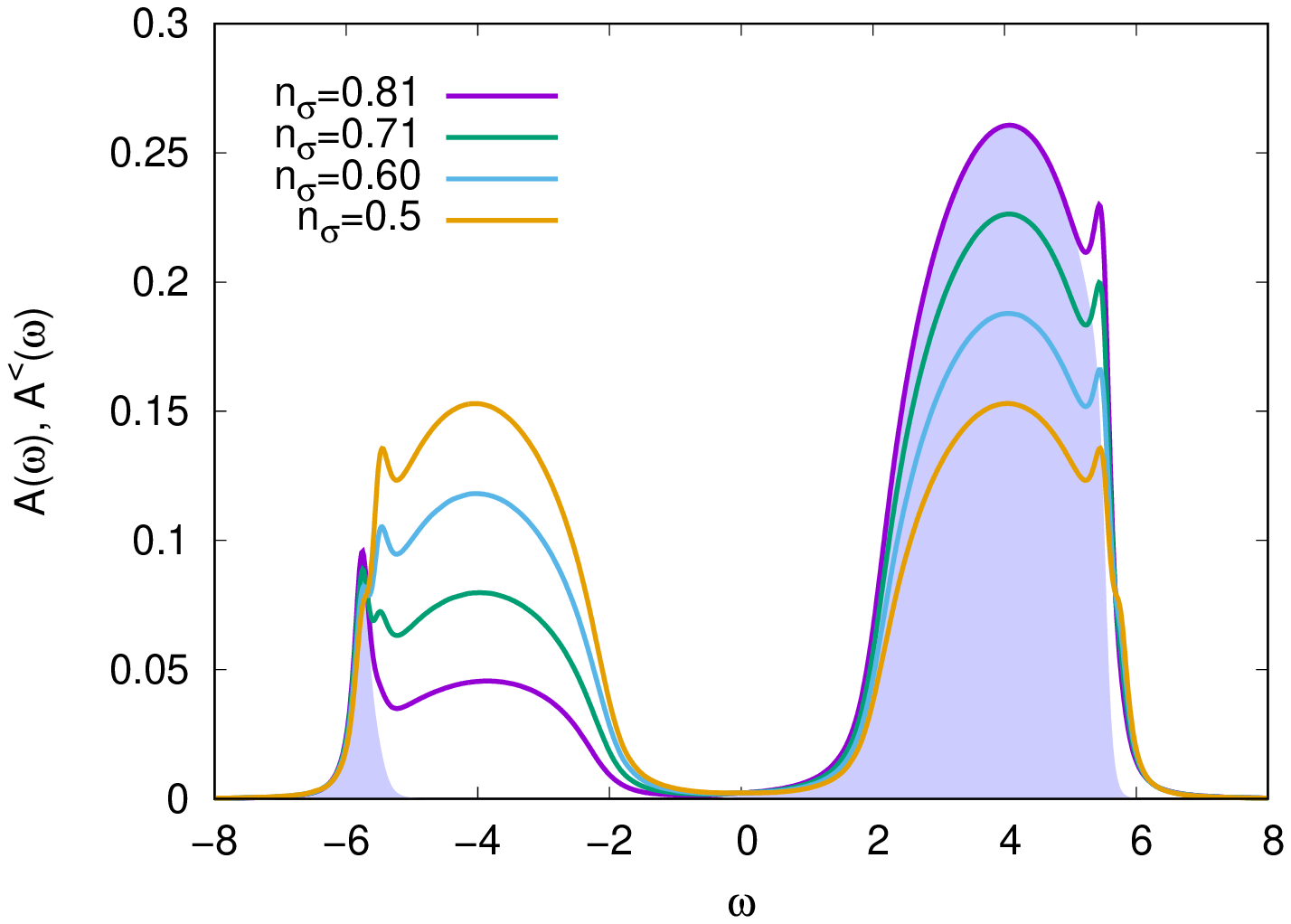}
\caption{OCA spectral functions $A(\omega)$ (lines) and occupations $A^<(\omega)$ (shading, for the largest asymmetry only) for $\eta$-pairing states with $\beta_d=\beta_h=10$. The effective chemical potentials are $\mu_d=-\mu_h=3.5$ (left panel), $\mu_d=-\mu_h=4.5$ (middle panel) and $\mu_d=-\mu_h=5.5$ (right panel), while the fillings $n_\sigma$ are indicated in the labels. 
}
\label{fig_spectra}
\end{center}
\end{figure*}

We find that the order parameter is remarkably robust against changes in the total filling and to a doublon-holon imbalance. 
For $\mu_d=-\mu_h=5.5$, which in the particle-hole symmetric case corresponds to $n_d=0.46$, the $\eta$-pairing state survives up to $n_\sigma \approx 0.87$, $n_d-n_h\approx 0.78$ and a relative doublon-holon imbalance of $(n_d-n_h)/(n_d+n_h)\approx 0.82$. This indicates that $\eta$-pairing becomes possible as soon as long-lived doublons and holons coexist in the system, with a minority charge carrier population $\gtrsim 20\%$. Even at lower photo-doping levels, the order parameter shows little sensitivity to the total filling, or chemical doping. In the left panel of Fig.~\ref{fig_op} we indicate the chemical doping range up to 15\% (corresponding to the underdoped regime in cuprates) by a gray shading. In this regime, the order parameter remains comparable to the value in the particle-hole symmetric limit, which always maximizes the $\eta$-pairing order. 

Figure~\ref{fig_spectra} plots the OCA spectral functions corresponding to three choices of $\mu_d=-\mu_h$ and the indicated total fillings $n_\sigma$ ($\beta_d=\beta_h=10$). In these figures, the lines plot the spectral functions $A(\omega)$, while the shaded regions indicate the occupations $A^<(\omega)$ for the most strongly doped systems. All shown results correspond to superconducting states. In the spectra for the intermediate photo-doping level, we notice the gap opening near the energies corresponding to the effective chemical potentials $\mu_d=-\mu_h=4.5$, and small quasi-particle peaks near the gap edges. The superconducting gap becomes more pronounced at lower effective doublon/holon temperatures. However, at very low effective temperatures, the DMFT solution with low-order strong-coupling solvers produces causality-violating artifacts in the energy-dependent spectral functions \cite{Geng2026}, which may not immediately destabilize the Bethe lattice calculation with the simplified self-consistency relation \eqref{eq_self}, but nevertheless results in unreliable spectra. To avoid such complications, we only present spectra for $\beta_h=\beta_d=10$.

Note that without photo-doping, equilibrium systems with $U>U_c$, $n_\sigma>0.5$ and low temperature exhibit a narrow quasi-particle band near the edge of the upper Hubbard band, or even inside the gap \cite{Georges1996,Werner2007}. The nonequilibrium simulations show that the corresponding heavily renormalized Fermi-liquid state is quickly destroyed by the introduction of doublons and holons, so that the photo-doped spectral functions look like those of a (high-temperature) Mott insulator with an asymmetric partial population of the upper and lower Hubbard bands. The persistence of a Mott gap in the photo-doped state is crucial for $\eta$ pairing. This gap protects the system from rapid heating through doublon-holon recombination processes, at least if the gap is larger than the width of the Hubbard bands \cite{Sensarma2010,Eckstein2011,Murakami2025}.  

\begin{figure}[h]
\begin{center}
\includegraphics[angle=-90, width=\columnwidth]{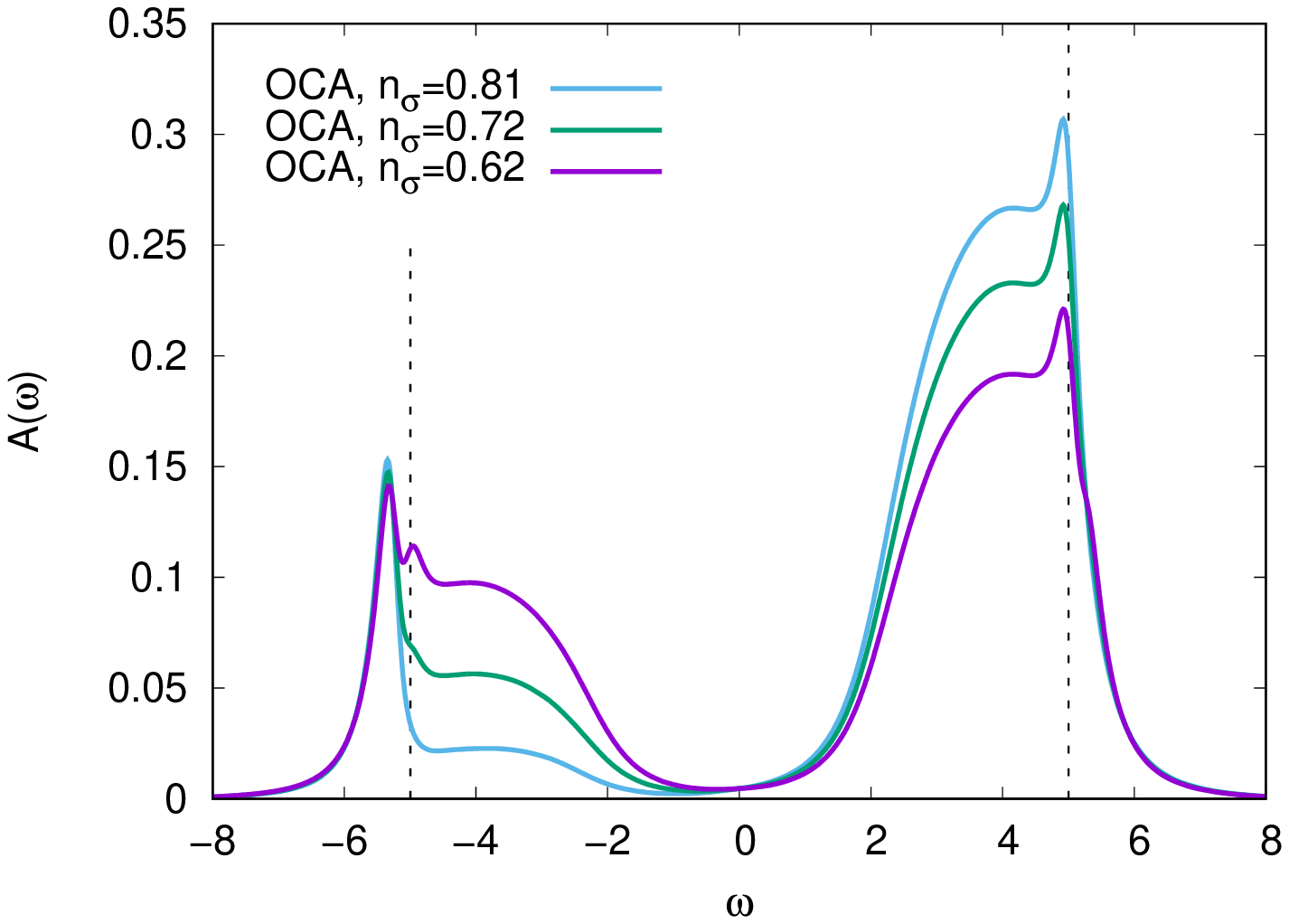}
\includegraphics[angle=-90, width=\columnwidth]{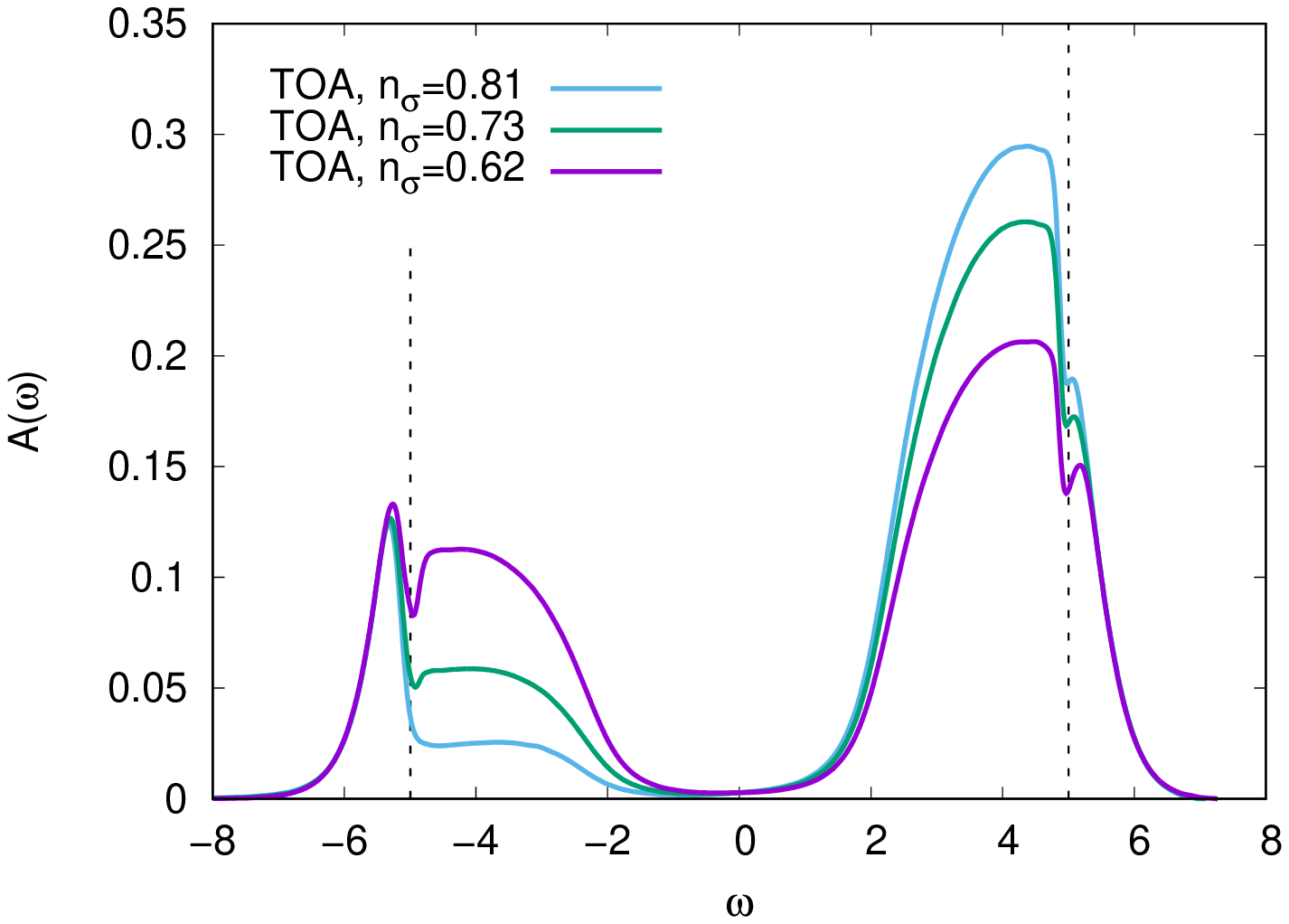}
\caption{
OCA and TOA spectral functions for $\beta_d=\beta_h=10$, $\mu_d=-\mu_h=5.0$ (vertical dashed lines) and three different fillings $n_\sigma$. 
}
\label{fig_spectra_TOA}
\end{center}
\end{figure}

As noted in Ref.~\onlinecite{Geng2026}, the spectral functions converge more slowly with increasing diagram order of the strong-coupling impurity solver than the $\eta$-pairing order parameter, so that there are significant corrections from the third order diagrams. This becomes even more pronounced in the particle-hole asymmetric solutions considered here. The OCA and TOA spectra for the calculation with $\mu_d=-\mu_h=5$ and three different fillings are shown in Fig.~\ref{fig_spectra_TOA}. While the order parameters predicted by these two solvers are consistent (see Fig.~\ref{fig_op}), the spectra show qualitative differences. For example, the gap features near $\mu_d$ and $\mu_h$, which are lacking in some of the OCA spectra, are systematically present at the TOA level. Impurity solvers treating diagrams up to fourth order or even higher would be needed to fully converge the spectra.  

\subsection{Nonequilibrium steady-states with different effective doublon and holon temperatures}

In a generic particle-hole asymmetric photo-doping setup, the effective doublon and holon temperatures can be different. Although these charge carriers are expected to quickly reach a common effective temperature in the conventional Hubbard model, since they scatter off each other, long-lived states with different doublon and holon temperatures are a characteristic feature of models with particle-hole asymmetric boson coupling, such as Hirsch's dynamic Hubbard model \cite{Hirsch2001,Werner2016}. It is thus interesting to explore how the $\eta$-pairing order parameter reacts to such an imbalance in effective temperatures. 

To study this problem in the standard Hubbard model, we impose nonequilibrium distribution functions with symmetric effective chemical potentials $\mu_d=-\mu_h$, similar to $f_2(\omega)$ (Fig.~\ref{fig_f}), but with separate effective doublon and holon temperatures $\beta_d\ne\beta_h$. In this setup, the doublon and holon densities are approximately the same, which allows us to clarify the effect of the temperature asymmetry. More explicitly, the distribution function imposed on the normal $\uparrow\uparrow$ element in the Nambu basis is chosen as
\begin{equation}
f_{\uparrow\uparrow}(\omega)
=
\left[1-f_{\bar\beta}(\omega)\right]
f_{\beta_d}(\omega-\mu_d)
+
f_{\bar\beta}(\omega)
f_{\beta_h}(\omega-\mu_h),
\label{eq:two_temperature_distribution}
\end{equation}
where $\bar\beta =(\beta_d+\beta_h)/2$ and $f_\beta(\omega)=1/(e^{\beta\omega}+1)$ is the Fermi distribution function for inverse temperature $\beta$.
The distribution of the $\downarrow\downarrow$ element is then fixed by the Nambu relation
\begin{equation}
f_{\downarrow\downarrow}(\omega)=1-f_{\uparrow\uparrow}(-\omega).\label{eq_downdown}
\end{equation}
A subtle issue is the choice of the effective distribution for the anomalous components of the Nambu Green's function \eqref{eq_nambu}. In the particle-hole symmetric case, it is natural to impose the same effective temperature for the normal and anomalous components. This is also consistent with the findings from real-time simulations with electron-phonon coupling \cite{Ray2025}, which showed that the effective temperatures extracted from the normal and anomalous components in the photo-induced $\eta$-pairing state are the same. However, if different effective doublon and holon temperatures are imposed in the normal components by fixing $f_{\uparrow\uparrow}(\omega)$ and $f_{\downarrow\downarrow}(\omega)$, the appropriate choice for the anomalous distribution is not a priori clear.

We considered two prescriptions. In the first prescription, no external distribution function is imposed on the anomalous components. Instead, only the diagonal distributions are fixed by Eqs.~\eqref{eq:two_temperature_distribution} and \eqref{eq_downdown}, while the anomalous lesser component is determined self-consistently by the DMFT loop. Unfortunately, this turns out to be numerically ill-behaved. The problem originates from the fact that both the retarded and lesser components of the anomalous spectral functions change sign near $\mu_d$ and $\mu_h$. If the zero crossings of the two functions do not exactly coincide, this leads to unphysical pole structures in $f_{\uparrow\downarrow}(\omega)$ and $f_{\downarrow\uparrow}(\omega)$. In the second prescription, we introduce a separate effective inverse temperature $\beta_\text{an}$ for the anomalous components, in the range $\text{min}(\beta_d,\beta_h)\le \beta_\text{an}\le \text{max}(\beta_d,\beta_h)$, and study the effect of $\beta_d$, $\beta_h$ and $\beta_\text{an}$ on the $\eta$-pairing order parameter in the photo-doped state. Note that since the anomalous spectra are antisymmetric, the effective anomalous temperatures on the electron and hole-doped side must be the same. 

\begin{figure*}[t]
\begin{center}
\includegraphics[angle=-90, width=0.65\columnwidth]{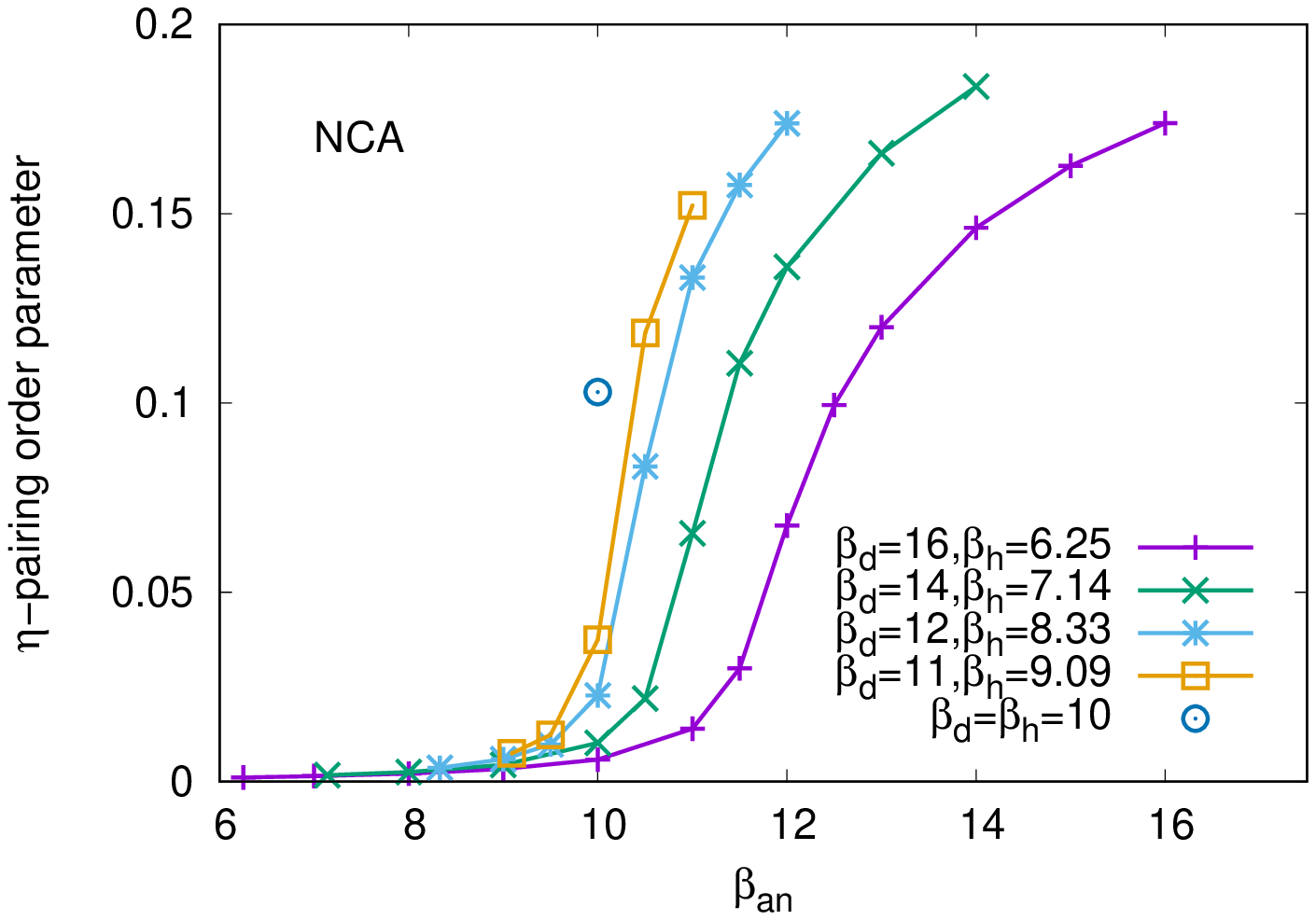}\hfill
\includegraphics[angle=-90, width=0.65\columnwidth]{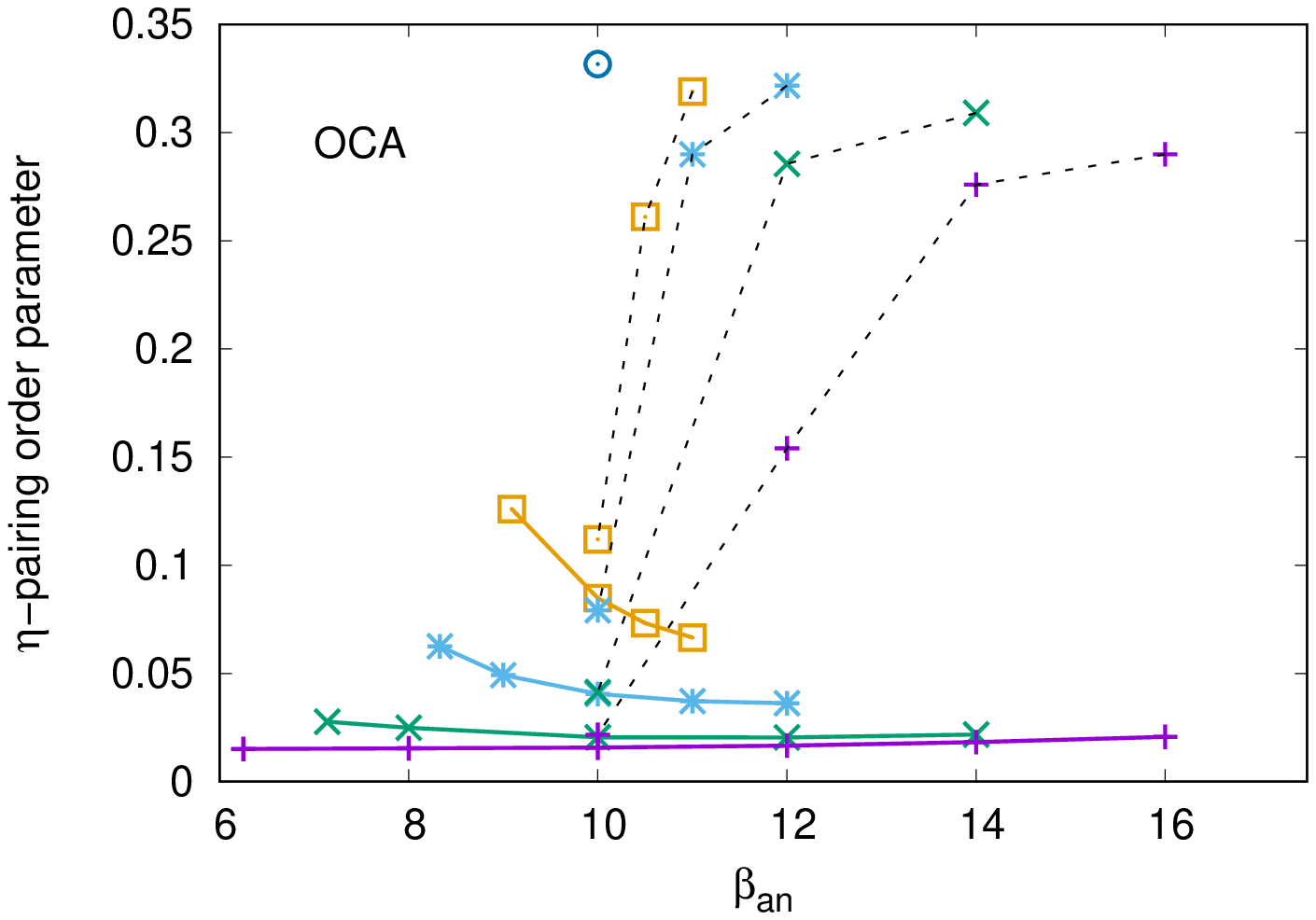}\hfill
\includegraphics[angle=-90, width=0.62\columnwidth]{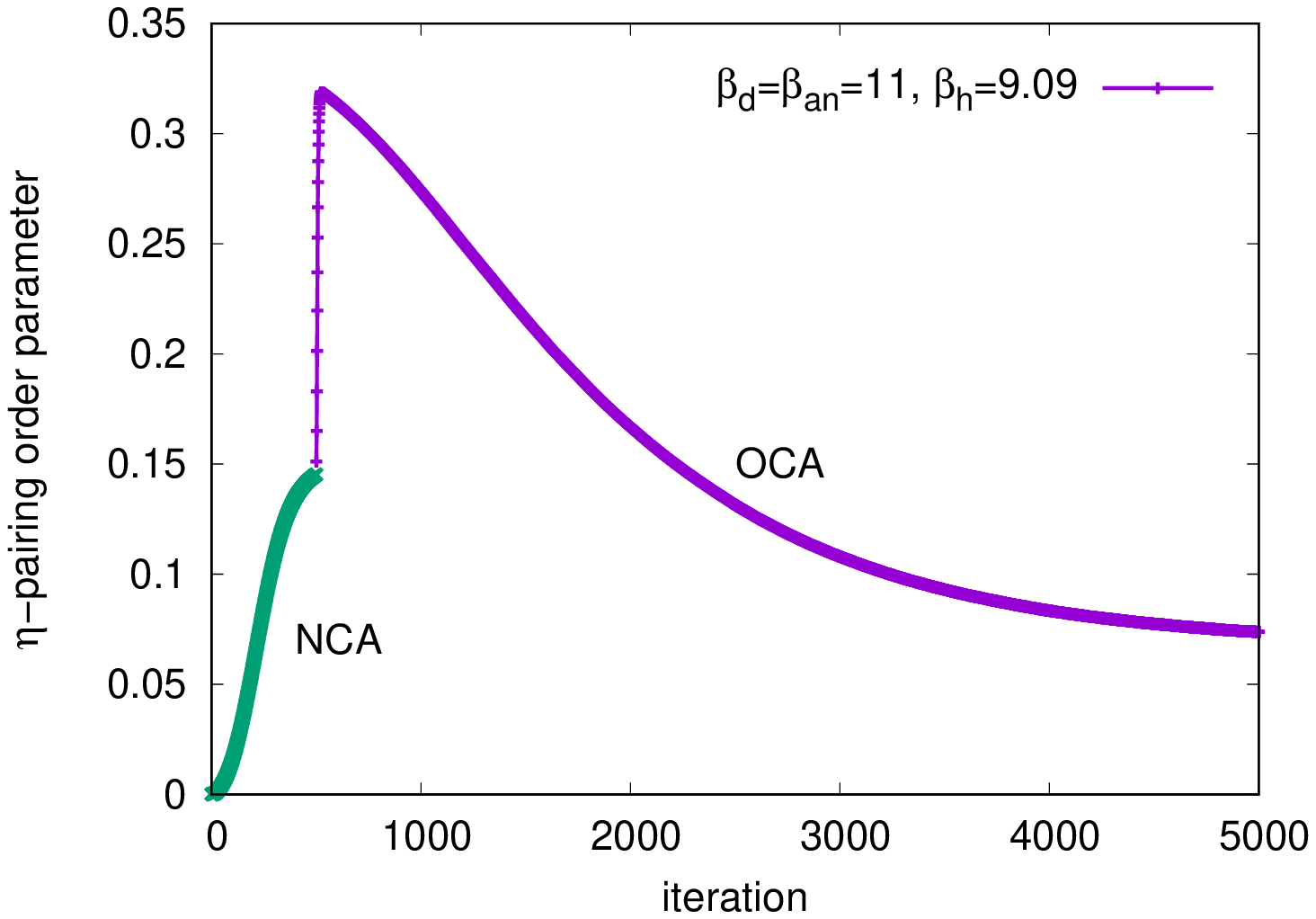}
\caption{$\eta$-pairing order parameter for a system with the indicated doublon and holon inverse temperatures, plotted as a function of the effective inverse temperature of the anomalous components ($U=8$, $\mu_d=-\mu_h=5.0$). The left panel shows NCA results and the middle panel OCA results. In the OCA panel, points connected by solid lines show the converged results, and points connected by dashed black lines the maximum order parameter reached in the DMFT self-consistency loop. The right panel shows an example of the evolution of the order parameter during the DMFT iterations. The first 500 iterations are performed at the NCA level, before switching to OCA.
}
\label{fig_2beta}
\end{center}
\end{figure*}

Figure~\ref{fig_2beta} plots the $\eta$-pairing order parameters for different sets of $(\beta_d,\beta_h)$ as a function of $\beta_\text{an}$. As reference system (blue dot) we choose a photo-doped particle-hole symmetric model with $U=8$, $\beta_d=\beta_h\equiv\beta_\text{ref}=10$ and $\mu_d=-\mu_h=5$. In the particle-hole asymmetric cases, we choose $\beta_d$ and $\beta_h$ such that the geometric mean reproduces this reference temperature, $\sqrt{\beta_d\beta_h}=\beta_\text{ref}$, keeping the effective chemical potentials unchanged. At the NCA level (left panel), the steady-state DMFT solutions for these imposed effective temperatures show that for $\beta_\text{an}=\beta_\text{ref}=10$, the $\eta$-pairing order parameter decreases rapidly with increasing asymmetry in the doublon and holon effective temperatures. However, if $\beta_\text{an}$ is closer to the lower of the two effective temperatures (in our setup the doublon effective temperature), then the asymmetric setup can yield an enhanced order parameter. 

The suppression of the order parameter with temperature asymmetry at $\beta_\text{an}=\beta_\text{ref}$ is also found at the OCA level (middle panel), although the $\beta_\text{an}$-dependence of the converged solution (solid lines) is qualitatively different from the NCA result. In particular, the converged order parameter for $\beta_\text{an}=\beta_d$ is much lower. An inspection of the convergence behavior with DMFT iterations (right panel) reveals an unusual behavior. After the (almost) convergence of the $\eta$-pairing order parameter at the NCA level at iteration 500, we switch on the OCA correction. After this, the order parameter initially more than doubles from $\approx 0.15$ to $0.32$, before decaying to a value which is smaller than the NCA order parameter. If we plot the order parameters corresponding to the transient maxima (symbols connected by black dashed lines in the middle panel), the curves look qualitatively similar to the NCA results. This behavior suggest that the true self-consistent solution of the DMFT equations has an anomalous distribution which is different from a Fermi function in the energy range of the Hubbard bands. 

We can try to extract  a rough estimate of $\beta_\text{an}$ from a fit to the anomalous spectra in the unconstrained simulation, although the aforementioned pole structures do not permit a proper definition of an effective temperature. These fits indicate that for a small asymmetry between $\beta_d$ and $\beta_h$, the effective anomalous inverse temperature $\beta_\text{an}$ tracks the lower of the two values, i.e. the hotter of the two effective temperatures. On the other hand, for large asymmetry, $\beta_\text{an}$ is closer to the average $(\beta_d+\beta_h)/2$. This means that in the case of the yellow curve in Fig.~\ref{fig_2beta}, the unconstrained solution would be near the lowest $\beta_\text{an}$ value, while in the case of the violet curve it would be closer to the middle point $\beta_\text{an}\approx 11$. In either case, the solutions for these imposed effective anomalous temperatures have a strongly suppressed order parameter. This shows that the $\eta$-pairing state is fragile against an asymmetry in the doublon and holon effective temperatures, at least if we impose thermal-like distributions for both the normal and anomalous components. 

\begin{figure*}[t]
\begin{center}
\includegraphics[angle=-90, width=0.32\textwidth]{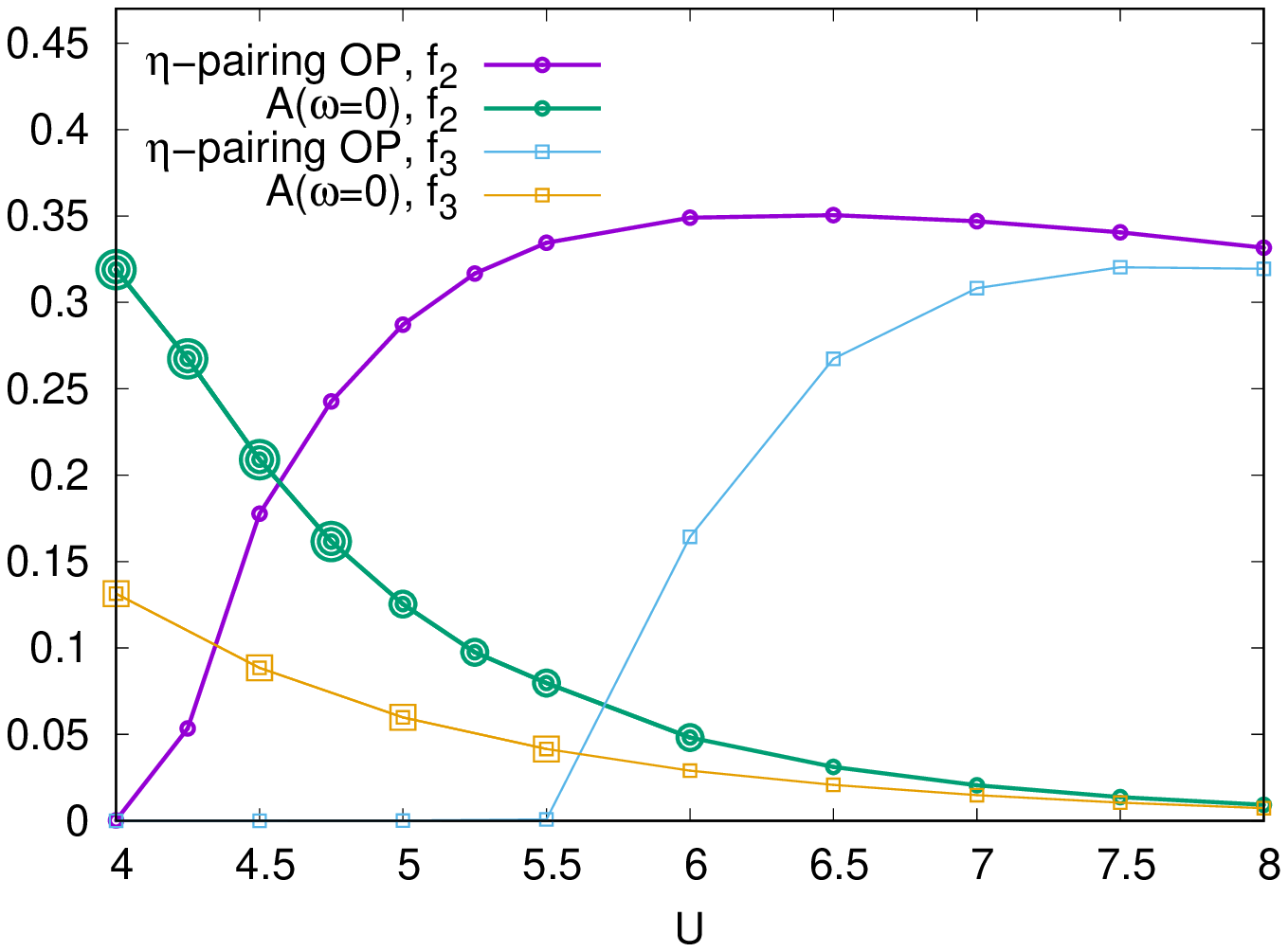}\hfill
\includegraphics[angle=-90, width=0.32\textwidth]{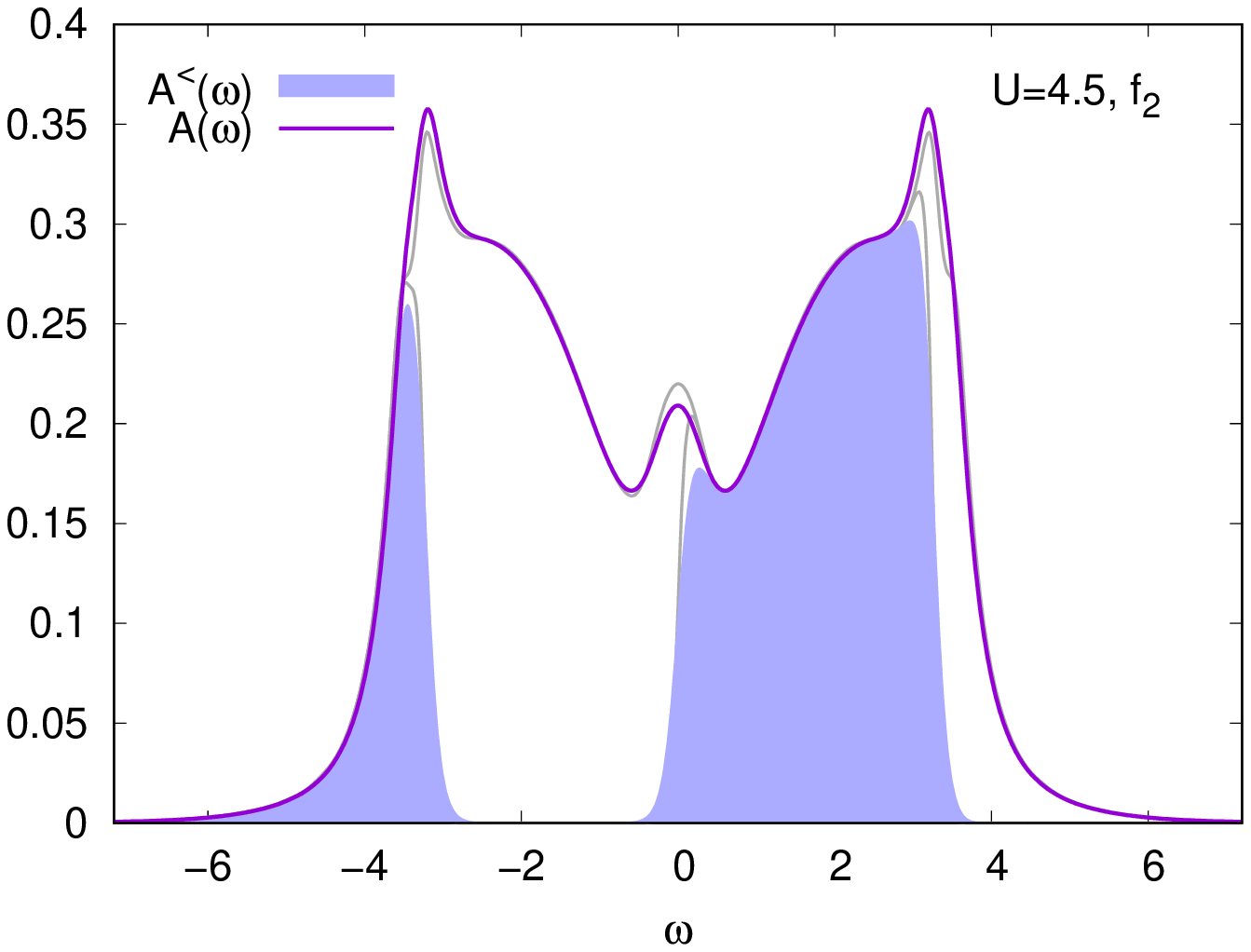}\hfill
\includegraphics[angle=-90, width=0.32\textwidth]{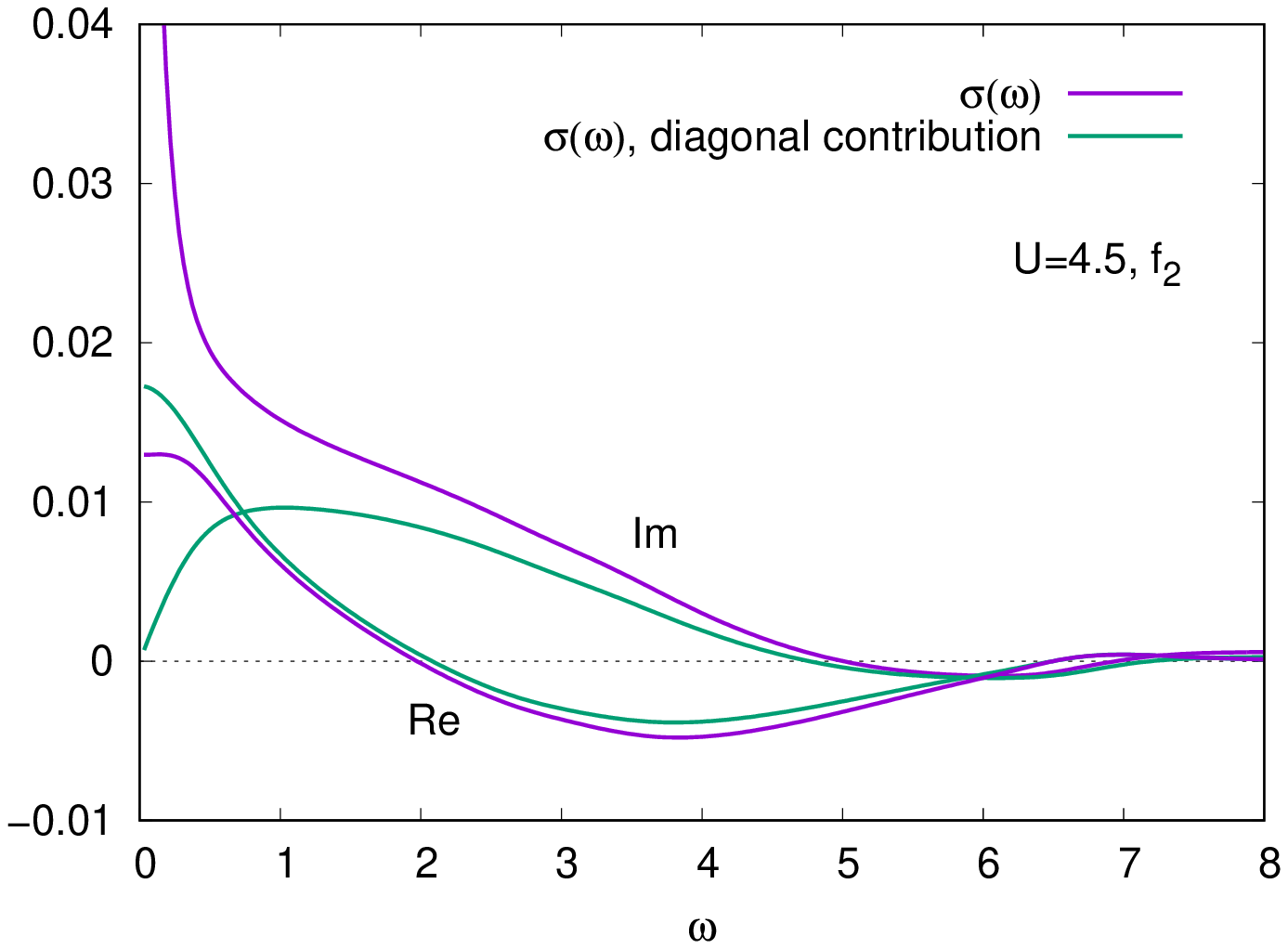}
\caption{Left panel: $U$-dependence of the $\eta$-pairing order parameter (OP) and density of states at $\omega=0$ for strongly photo-doped particle-hole symmetric systems with an $f_2$ type distribution (circles) and an $f_3$ type distribution (squares). Single circles and squares indicate a gapped spectrum, double circles and squares a pseudo-gapped spectrum, and triple circles a spectrum with a quasi-particle peak at $\omega=0$. (OCA results for $\beta_\text{eff}=10$.) Middle panel: OCA spectral function and occupation for $U=4.5$, $\mu_d=-\mu_h=3.75$, $\beta_\text{eff}=10$ and the $f_2$ distribution. Gray lines show $\beta_\text{eff}=20$ results. Right panel: Real and imaginary part of the optical conductivity $\sigma(\omega)$ for $U=4.5$, $\mu_d=-\mu_h=3.75$, $\beta_\text{eff}=10$ and the $f_2$ distribution (violet curves). For comparison, the green curves show the result obtained without the anomalous Green's function components. 
}
\label{fig_data_u}
\end{center}
\end{figure*}

\subsection{Photo-doped metals}

As mentioned in the introduction, all the experiments reporting light-induced superconducting-like states \cite{Fausti2011,Kaiser2014,Mitrano2016,Buzzi2020,Budden2021} involved correlated metals, not Mott insulators. Some of them (cuprates) are chemically doped Mott insulators, with $U$ larger than the critical $U_c$ for the bandwidth controlled Mott transition in the undoped compound. The realization of $\eta$-pairing in such systems has been discussed in Sec.~\ref{subsec_doped}. Here, we focus on half-filled metallic systems with $U<U_c$, which is a setup relevant for K$_3$C$_{60}$ \cite{Zadik2015} and $\kappa$-(BEDT-TTF)$_2$Cu[N(CN)$_2$]Br \cite{Jerome1991}. In the absence of a gap between the Hubbard bands, long-lived and effectively cold photo-doped doublon and holon populations are unlikely to exist. One can however imagine transient states with a partial doublon/holon population of the upper/lower Hubbard bands during laser driving, or on sub-picosecond timescales after an excitation pulse. It is thus interesting to explore if such photo-doped metal states can exhibit long-range order of the $\eta$-pairing type. 

In a first investigation, we compare the effects of an $f_2$ type distribution with Fermi edges in the Hubbard bands and an inverted (negative temperature) distribution near $\omega=0$ to those of an $f_3$ type distribution with a positive temperature in the low-energy region. We vary $U$ and shift $\mu_d=-\mu_h$ correspondingly, so that the doublon and holon densities remain approximately constant. The left panel of Fig.~\ref{fig_data_u} plots the OCA results for strongly photo-doped systems with $(U,\mu_d)=(8,5), (7.5,4.75), (7,4.5), \ldots$, $n_d=n_h\approx 0.4$ and $\beta_\text{eff}=10$. The two types of lines show the evolution of the $\eta$-pairing order parameter and the density of states at $\omega=0$. The nature of the spectral function is indicated by the symbols of the density of states curves: single circles and squares represent gapped spectra, double circles and squares pseudo-gapped spectra and triple circles spectra with a quasi-particle peak at $\omega=0$. For these strongly photo-doped systems, the solutions with the $f_3$ type distribution do not develop a quasi-particle peak down to $U=4$, and superconductivity is already absent in the pseudo-gapped states. In contrast, the solutions with $f_2$ type distribution exhibit $\eta$-pairing in the range $4.25\lesssim U \lesssim 4.75$, in which the spectral function exhibits a low-energy quasi-particle peak. The spectral function and occupation for $(U,\mu_d)=(4.5,3.75)$, $\beta_\text{eff}=10$ and the $f_2$ distribution are shown in the middle panel of Fig.~\ref{fig_data_u} (violet curves and shading). As the effective temperature is reduced from $\beta_\text{eff}=10$ to $\beta_\text{eff}=20$, the low-energy quasi-particle feature grows at the expense of the quasi-particle peaks near $\omega=\pm\mu_d$ (gray lines), and the $\eta$-pairing order parameter increases from 0.178 to 0.204. If we include the third order diagrams within TOA, the low-energy quasi-particle peak also grows, while the peaks near $\omega=\pm\mu_d$ turn into small dips, similar to the result shown in Fig.~\ref{fig_spectra_TOA}. The order parameter decreases to 0.127 at $\beta_\text{eff}=10$, but the solution remains superconducting. 

We also performed analogous $U$-scans for smaller $\mu_d=-\mu_h$ and hence lower photo-doping densities. The results for $n_d=n_h\approx 0.25$ remain qualitatively similar to those discussed above. For $n_d=n_h\approx 0.18$, we did not find any low energy quasi-particle feature down to $U=3$ at $\beta_\text{eff}=10$, while $\eta$-pairing is absent for $U\lesssim 3.5$. At lower effective temperatures, a superconducting state with population inversion in the low-energy quasi-particle band may however emerge. Note that $\beta_\text{eff}=10$ corresponds to a high temperature scale: in a system with a bandwidth of O(eV) it corresponds to an effective electronic temperature of O(1000)~K. 

At first sight, an inverted population in the quasi-particle band may seem unnatural. However, previous studies of electron-boson coupled systems with highly excited boson distributions \cite{Werner2016,Golez2017} revealed the transient stabilization of such population inverted electronic states. It is thus conceivable that in a photo-doped strongly correlated metal, energy released by doublon-holon recombination processes is transferred to the quasi-particle states, resulting in a  
negative temperature distribution within the quasi-particle band. Our results show that this type of photo-doped metal can exhibit $\eta$ pairing. 

A characteristic signature of superconductivity is a $\frac{1}{\omega}$-type divergence of the imaginary part of the optical conductivity $\sigma(\omega)$ at low frequencies, and the emergence of a dip or gap in the real part. These features have been clearly demonstrated for the $\eta$-pairing state in previous studies of the photo-doped Mott regime \cite{Li2020,Geng2026}. For completeness, we show in the right panel of Fig.~\ref{fig_data_u} the optical conductivity for the metallic $\eta$-pairing solution plotted in the middle panel. The conductivity has been calculated as described in the Ref.~\cite{Li2020}, and the green lines plot the result obtained without the anomalous components in the Green's function matrix. The presence of the anomalous components leads to the expected $\frac{1}{\omega}$-divergence in $\text{Im}\,\sigma(\omega)$ and a reduction of $\text{Re}\,\sigma(\omega)$ at low frequencies. Note that the $\eta$-pairing state can be understood in terms of a two-fluid picture, where condensed doublons and holons coexist with normal singlons \cite{Li2020}.
 
We now turn to metallic states with an $f_3$ type distribution and hence a positive temperature within the quasi-particle band. Here, we consider the model with $U<W=4$, but with strong enough correlations that the spectral function exhibits Hubbard satellites. The smooth switching between the three Fermi-like distributions in $f_3$ is imposed at $\omega=\pm \mu_\text{switch}$, which we choose near the lower edge of the Hubbard bands. Specifically, we use a $1-f_{\beta_\text{switch}}(\omega\pm\mu_\text{switch})$ switching function with $\beta_\text{switch}=\beta_\text{eff}/2$. The resulting nonequilibrium distributions mimic transient states which are potentially realizable in photo-doped strongly correlated metals, if there exists a relaxation bottleneck between the Hubbard bands and the low-energy quasi-particle band.        

\begin{figure*}[t]
\begin{center}
\includegraphics[angle=-90, width=0.49\columnwidth]{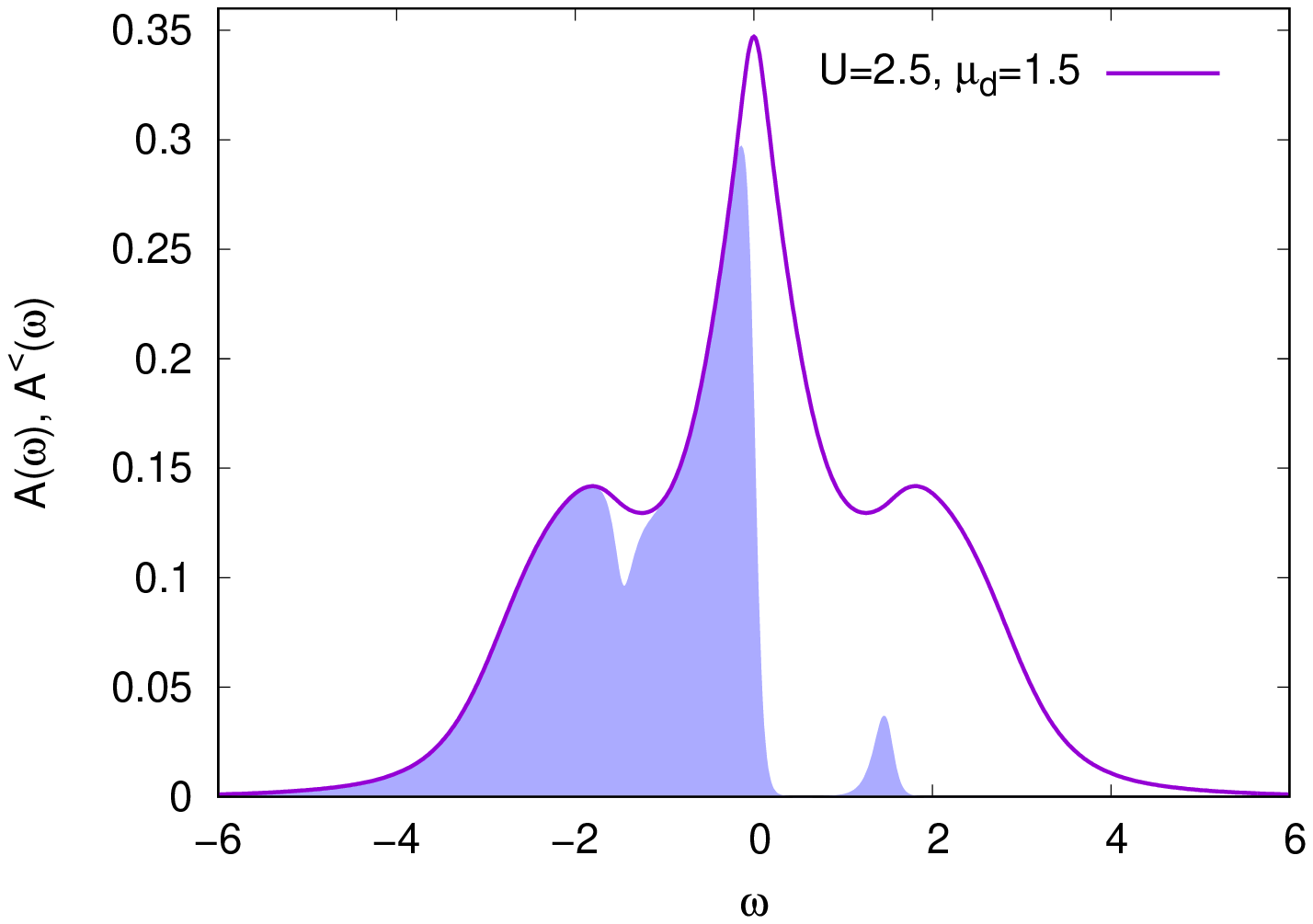}\hfill
\includegraphics[angle=-90, width=0.49\columnwidth]{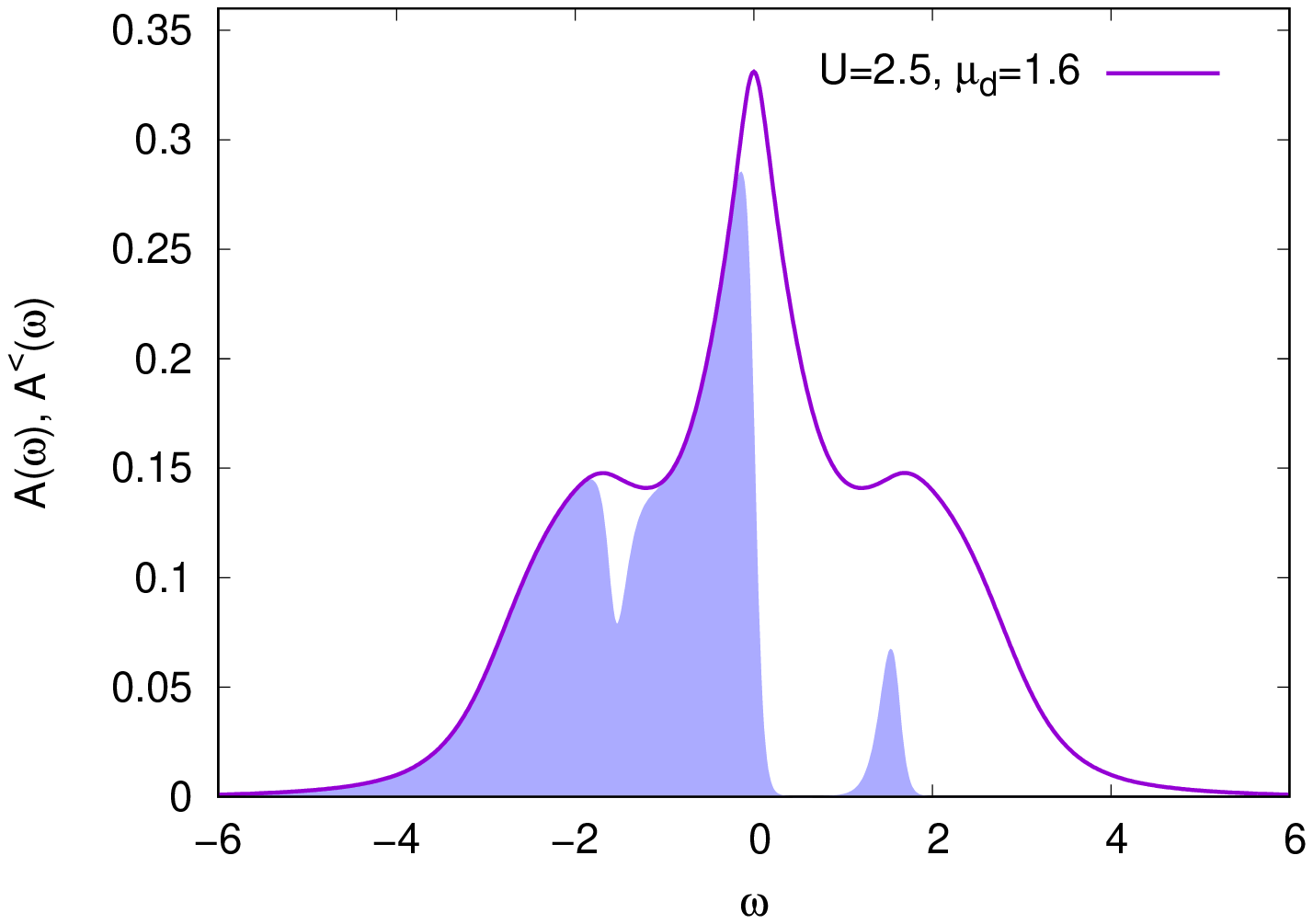}\hfill
\includegraphics[angle=-90, width=0.49\columnwidth]{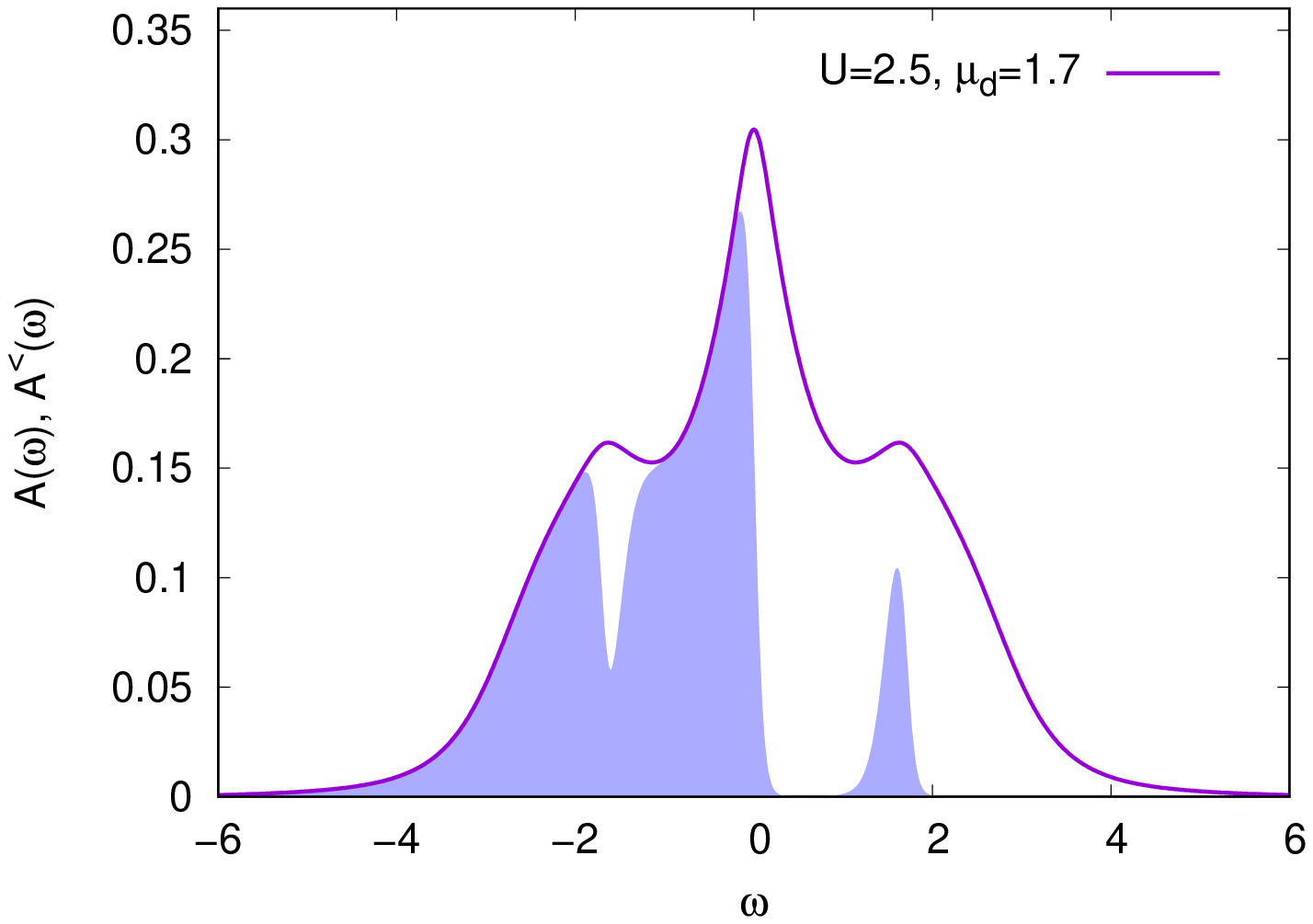}\hfill
\includegraphics[angle=-90, width=0.49\columnwidth]{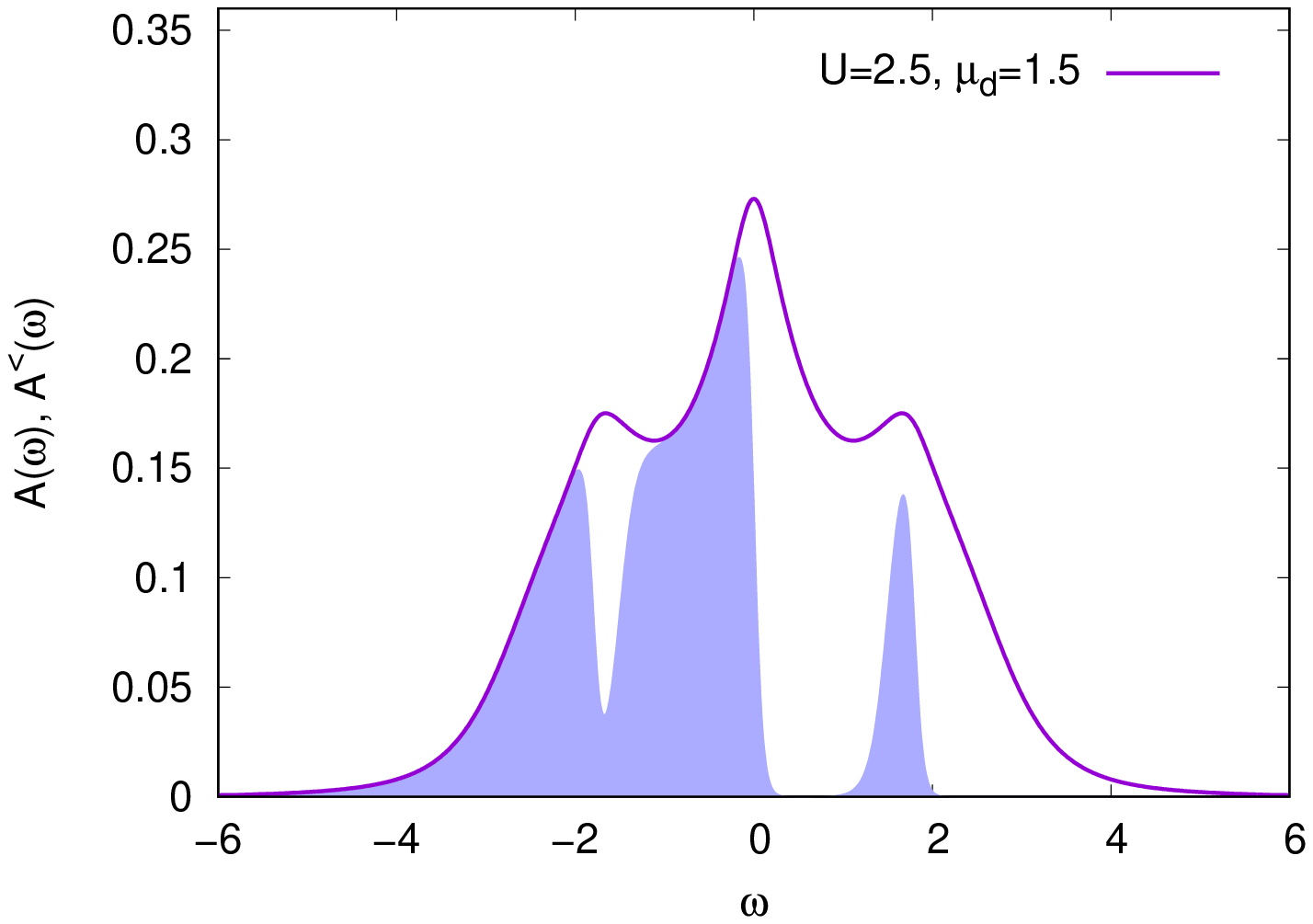}\\
\includegraphics[angle=-90, width=0.49\columnwidth]{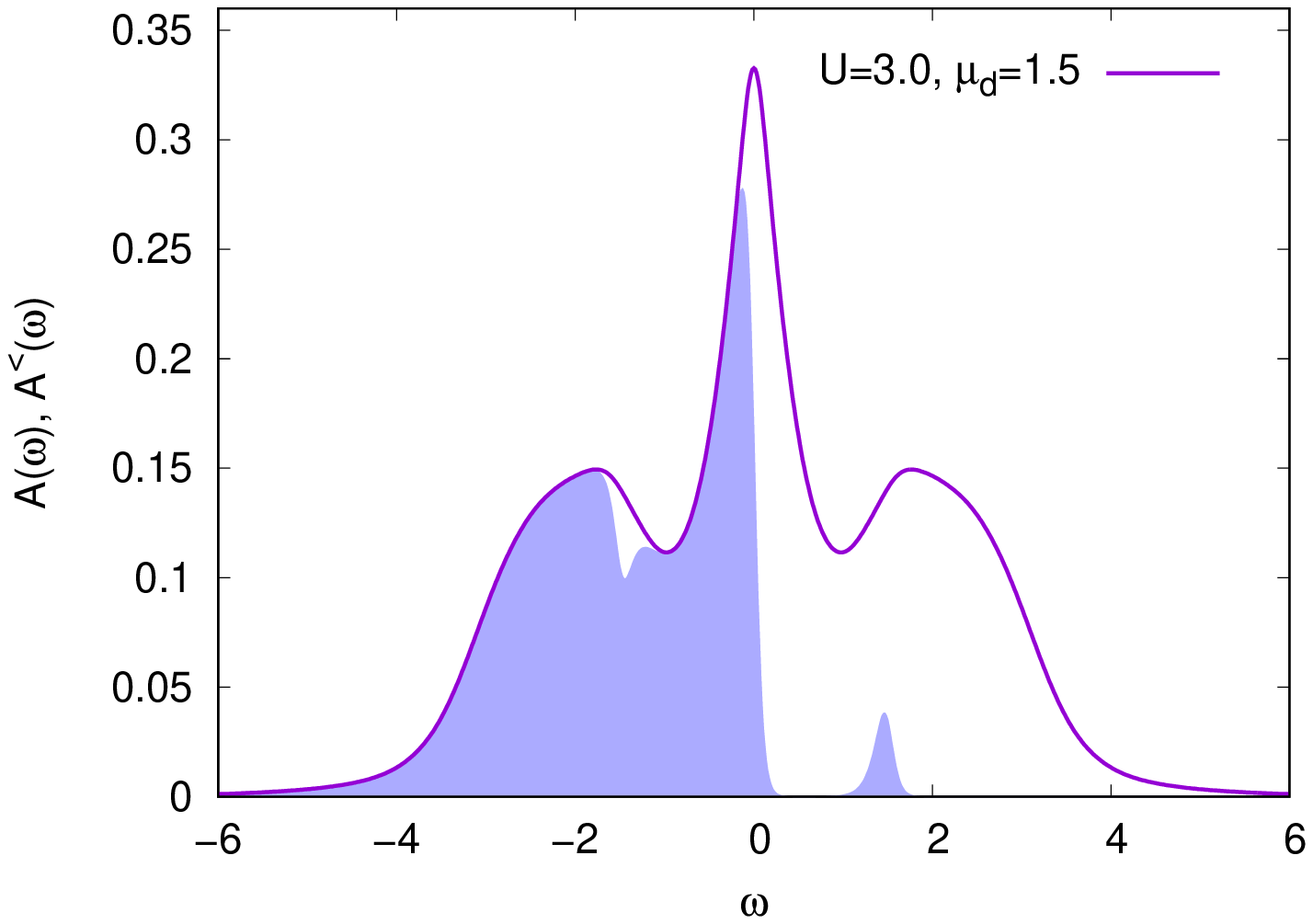}\hfill
\includegraphics[angle=-90, width=0.49\columnwidth]{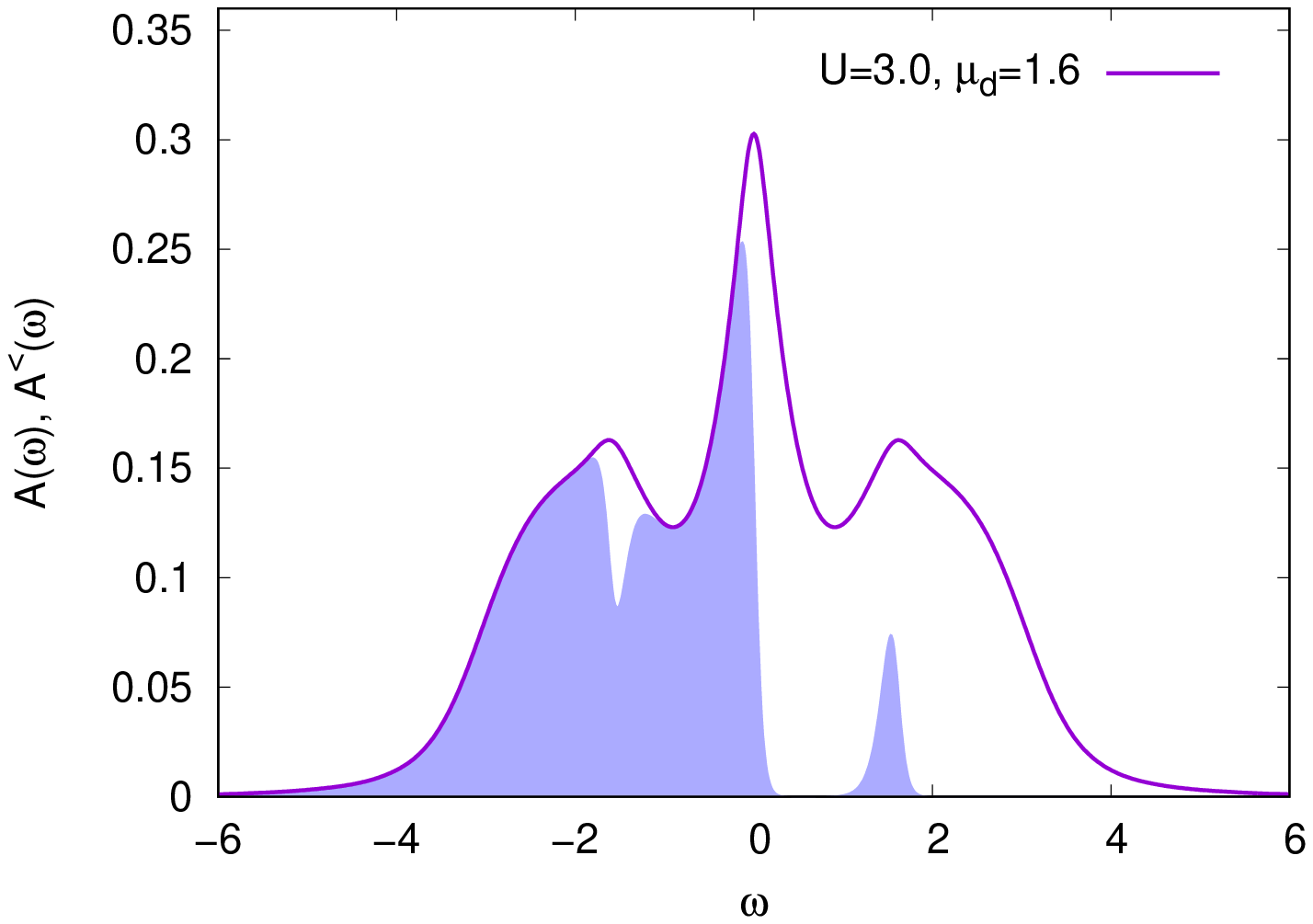}\hfill
\includegraphics[angle=-90, width=0.49\columnwidth]{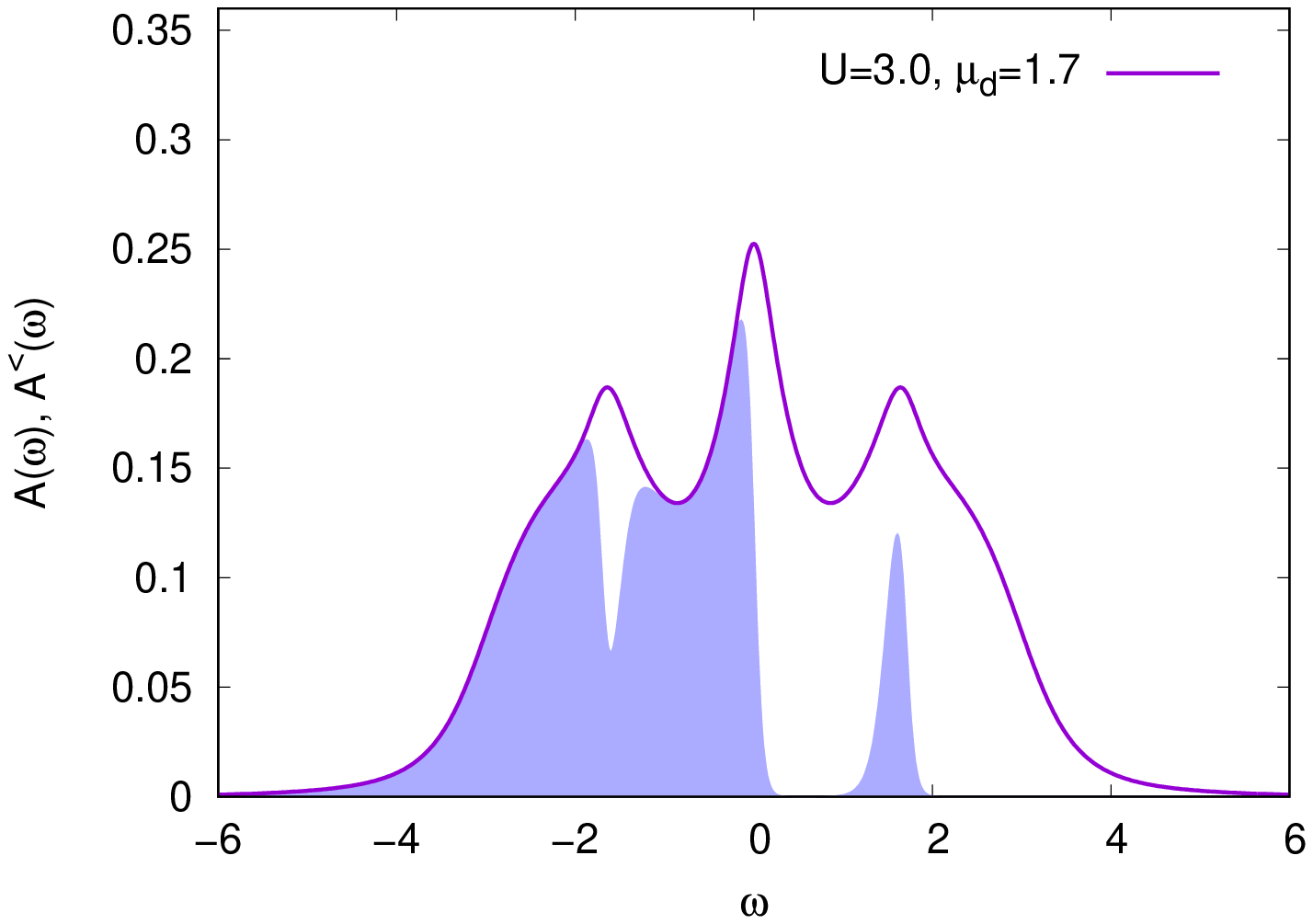}\hfill
\includegraphics[angle=-90, width=0.49\columnwidth]{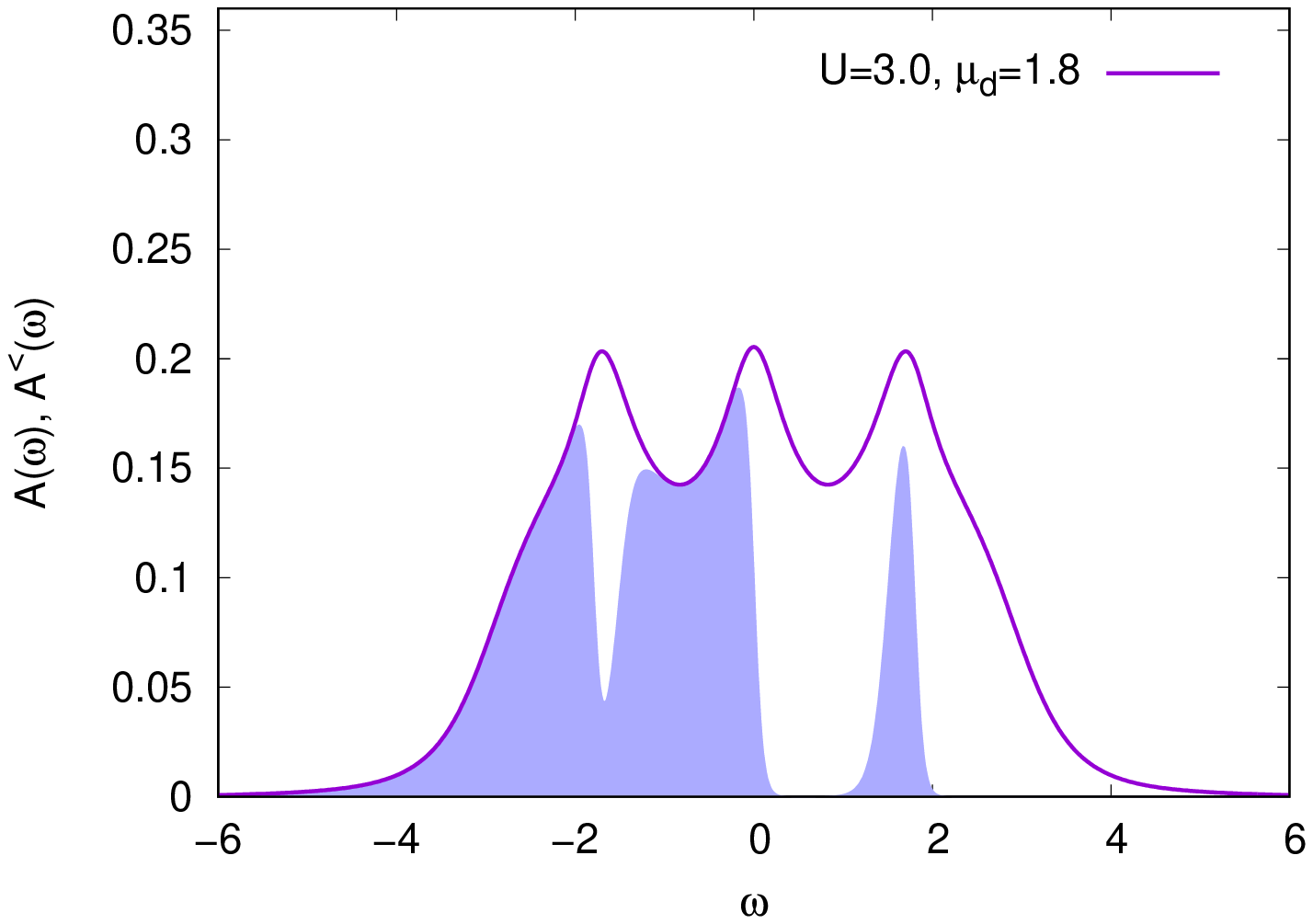}\\
\includegraphics[angle=-90, width=0.49\columnwidth]{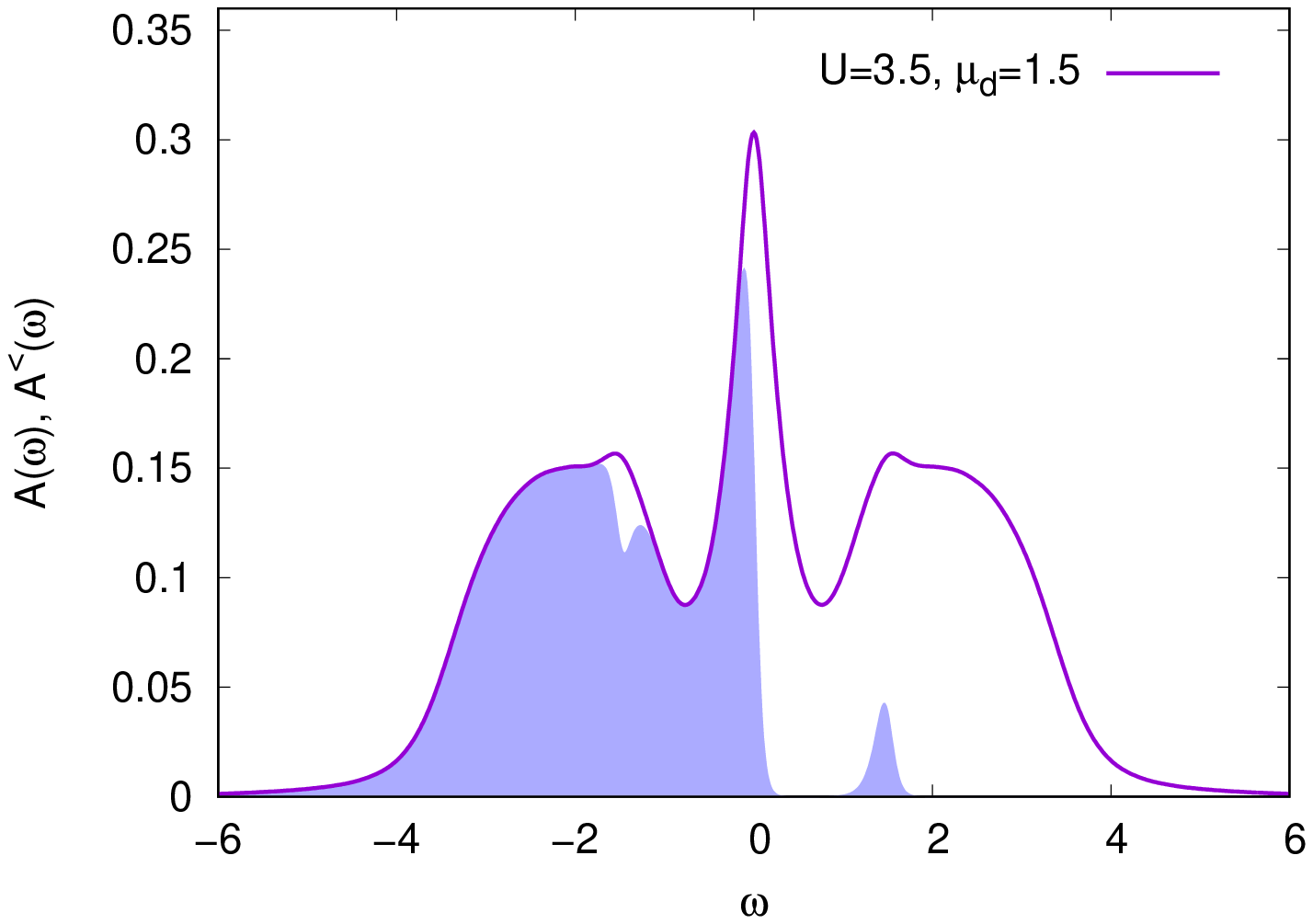}\hfill
\includegraphics[angle=-90, width=0.49\columnwidth]{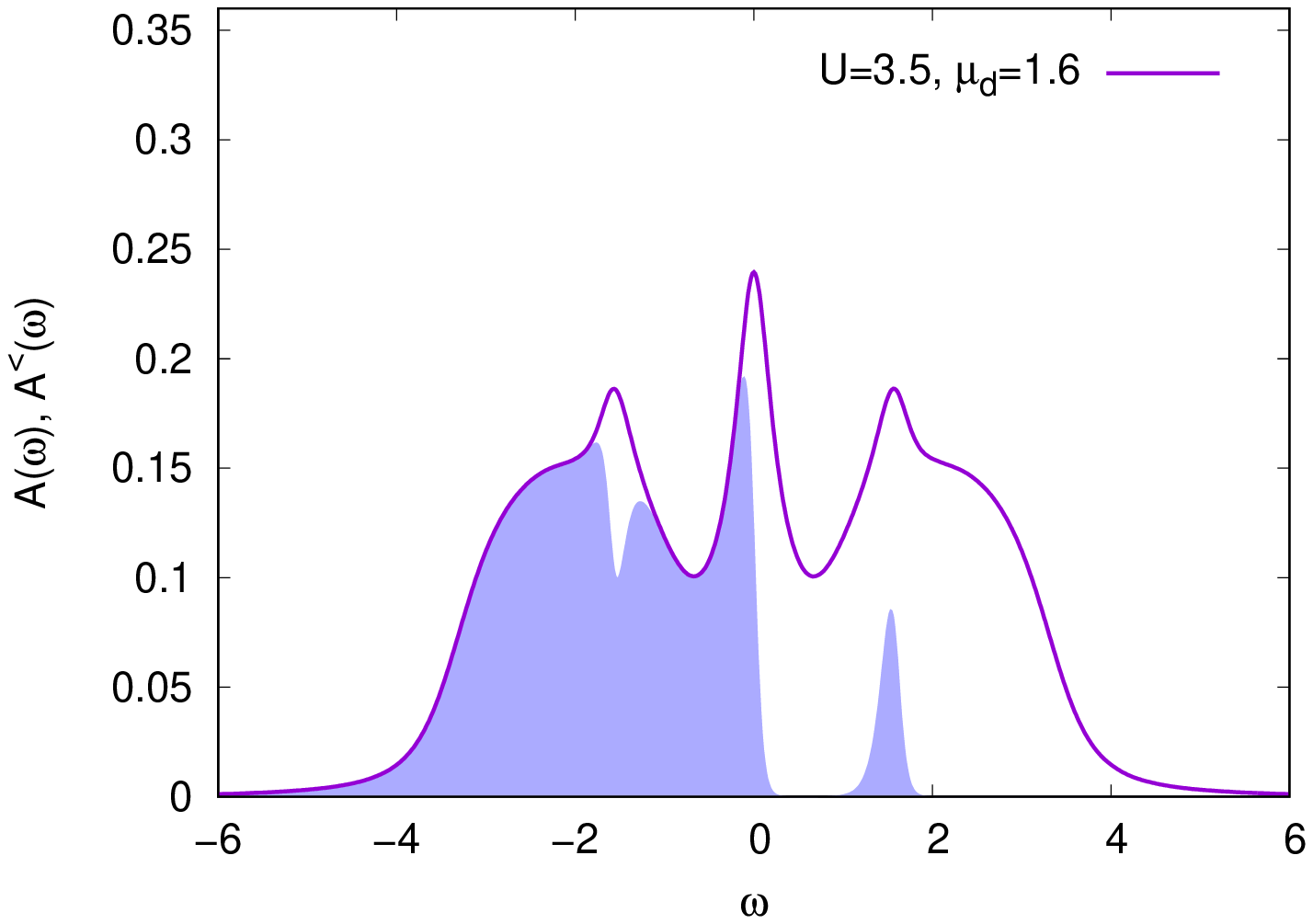}\hfill
\includegraphics[angle=-90, width=0.49\columnwidth]{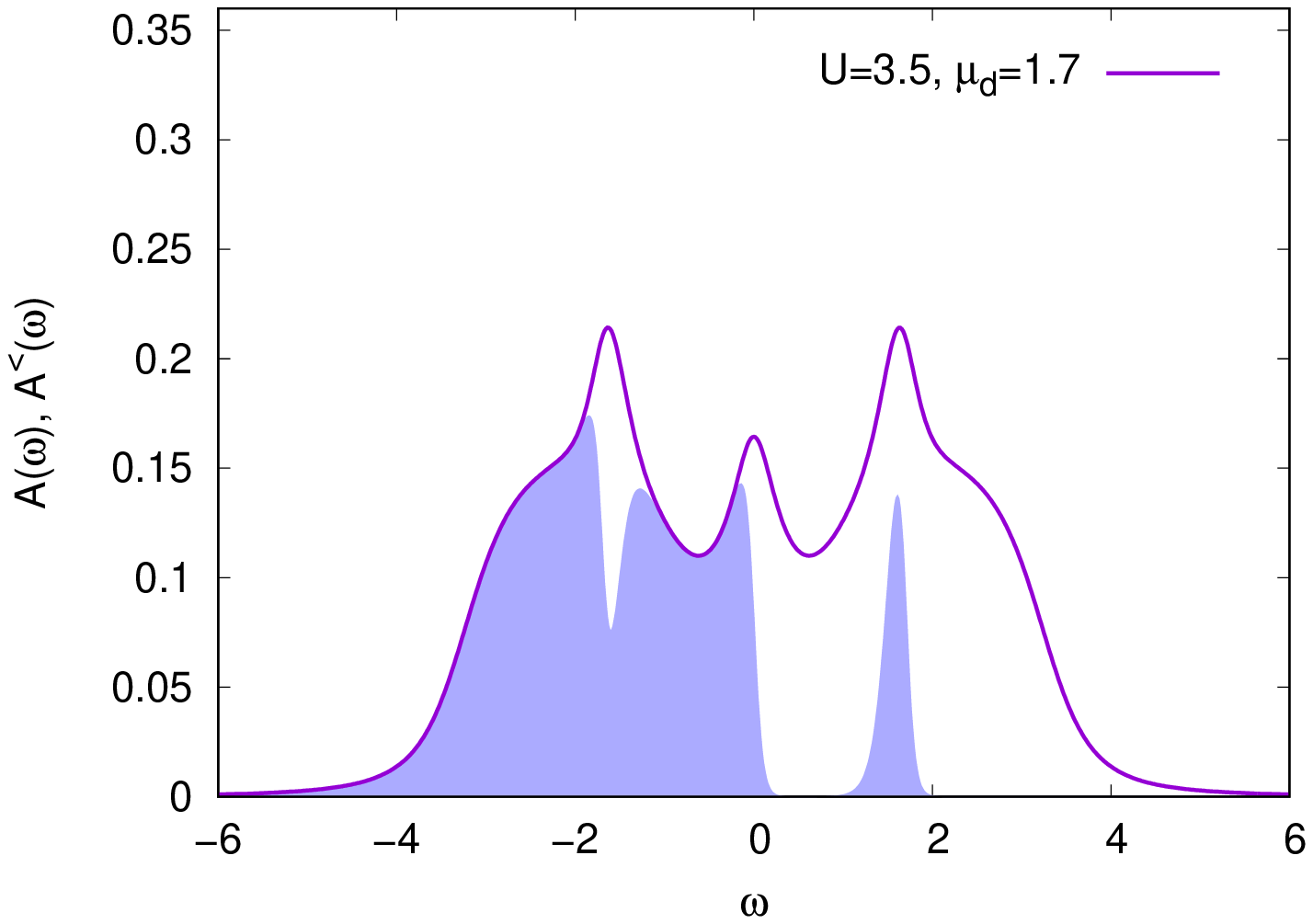}\hfill
\includegraphics[angle=-90, width=0.49\columnwidth]{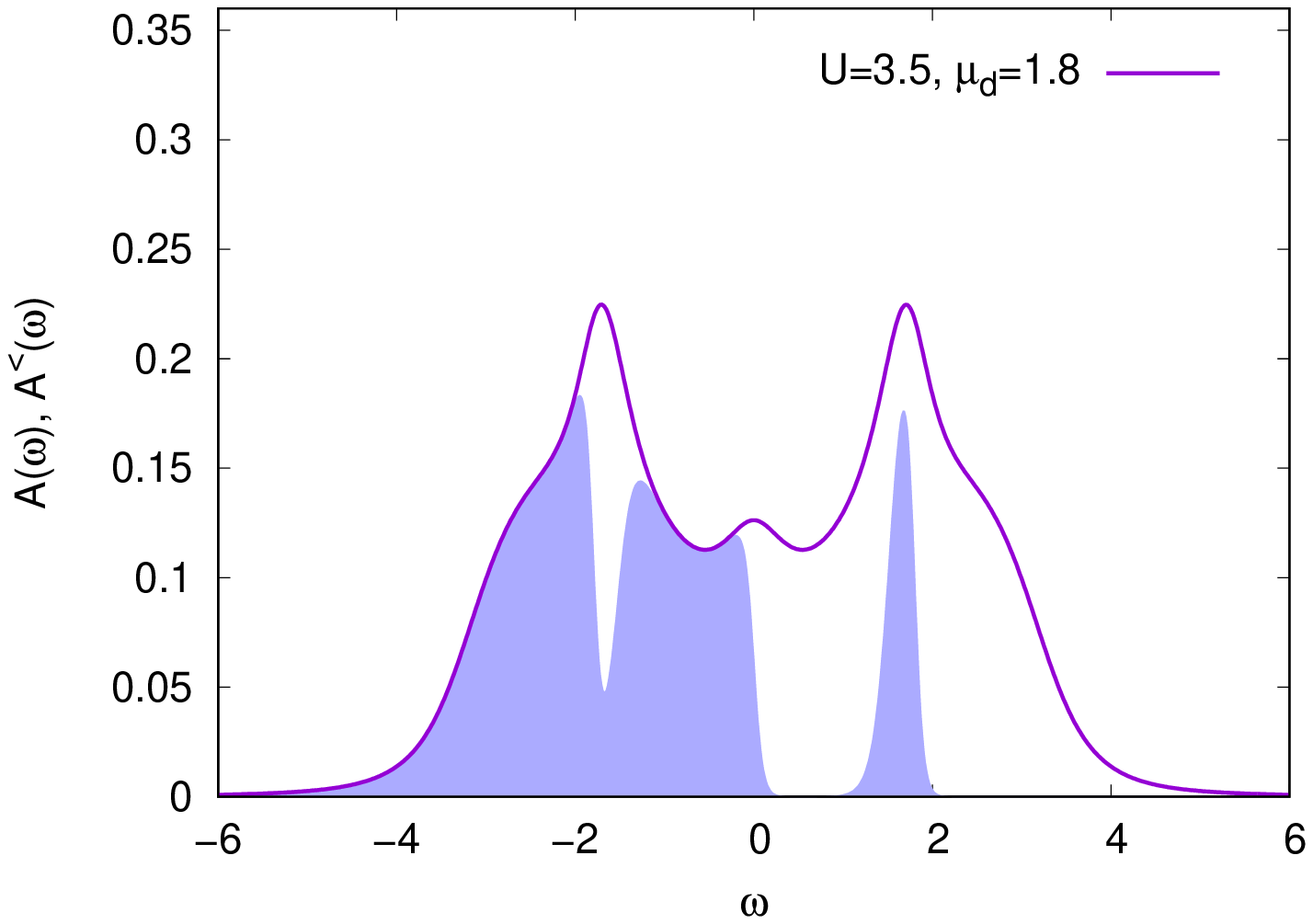}
\caption{OCA spectral functions $A(\omega)$ (lines) and occupations $A^<(\omega)$ (shading) for $U=2.5$, $3.0$, $3.5$ and the indicated values of $\mu_d=-\mu_h$. ($\mu_\text{switch}=1.5$,  $\beta_\text{eff}=20$ and $\beta_\text{switch}=10$.)}
\label{fig_metal}
\end{center}
\end{figure*}

Figure~\ref{fig_metal} plots the nonequilibrium spectral functions and occupations for $U=2.5$, $3.0$, $3.5$, $\mu_\text{switch}=1.5$ and a range of $\mu_d=-\mu_h$ corresponding to weakly and moderately photo-doped Hubbard bands. The effective inverse temperature for the doublons and holons is $\beta_d=\beta_h=20$ (same as the temperature distribution of the quasi-particle states). Increasing the photo-doping suppresses the coherent quasi-particle states near $\omega=0$ due to additional scattering channels. For $U\gtrsim 3.5$, close to the high-temperature end point of the Mott transition line \cite{Bluemer_thesis}, the system thus quickly ends up in a pseudo-gapped bad metal state without quasi-particles, reminiscent of a high-temperature Mott insulator. 

Despite the low effective doublon and holon temperature $\beta_\text{eff}=20$, which would be difficult to realize in an actual experiment or time-dependent photo-doping simulation of a metallic system, none of the setups shown in Fig.~\ref{fig_metal} exhibit $\eta$-pairing. We considered even lower effective temperatures, and also different values of $\mu_\text{switch}$ and $U$, but could not find any superconductivity in weak-to-moderately photo-doped metallic systems with $f_3$ type distribution. The only nonequilibrium steady states which showed $\eta$ pairing were strongly photo-doped systems near the metal-insulator transition, which are however better characterized as photo-doped narrow-gap Mott insulators. In such systems, even with a three-Fermi-edge type nonequilibrium distribution function, superconductivity can develop. This shows that the details of the population in the gapped region does not matter as far as $\eta$-pairing is concerned. It is however unlikely that a reasonably long-lived transient state with a three-Fermi-edge distribution exists in a strongly photo-doped correlated metal or narrow gap insulator. We therefore conclude that $\eta$-pairing does not generically occur in photo-excited metals with $U<U_c$ and a positive temperature distribution in the low-energy quasi-particle band.

\section{Conclusions}
\label{sec_conclusions}

Numerical simulations of photo-doped single-band Hubbard models revealed $\eta$-pairing superconductivity with very high effective critical temperature \cite{Werner2019b,Li2020,Geng2026}. These results were obtained for particle-hole symmetric systems in the large-gap Mott regime. In this work, we investigated the robustness of the $\eta$-pairing states against changes in the doublon and holon densities, the total filling, as well as the doublon and holon temperatures. We found that for $U$ larger than the critical value $U_c$ of the bandwidth controlled Mott transition, $\eta$-pairing is remarkably robust against a doublon-holon imbalance or changes in the total filling. It can exist even in almost completely filled systems, if these host a sufficient density of holons (empty sites). While in equilibrium, chemically doped Mott insulators are metallic and may feature a strongly renormalized quasi-particle band at low temperature, these coherent quasi-particles are quickly destroyed by photo-doping. The spectra of the photo-doped metastable states thus exhibit a (partially filled-in) gap, and in this regard are more reminiscent of a high-temperature Mott insulator than a good metal. In such large-gap chemically doped and photo-doped Mott systems, $\eta$-pairing with high effective critical temperature can be generically realized. Our findings suggest that if a sufficiently long-lived and effectively cold photo-doped population can be induced in underdoped cuprates, for example using an entropy cooling protocol \cite{Werner2019a}, $\eta$-pairing may potentially emerge.

On the other hand, we found that an asymmetry in the effective doublon and holon temperatures, as expected for example in the photo-doped dynamic Hubbard model \cite{Hirsch2001,Werner2016} or other systems with particle-hole asymmetric electron-boson coupling, easily suppresses the $\eta$-pairing order parameter. Simulations without constraint on the distribution function of the anomalous component can sometimes converge to solutions with enhanced order. These however feature unphysical-looking nonequilibrium anomalous distributions with pole structures. To circumvent such potential numerical artifacts, we imposed the effective inverse temperature $\beta_\text{an}$ of the anomalous components and treated it as a separate parameter. In this setup, the calculations were stable and revealed that if  $\beta_\text{an}$ is fixed to the reference inverse temperature $\beta_\text{ref}$, while $\beta_d$ ($\beta_h$) are increased (decreased) relative to it, the $\eta$-pairing order parameter is suppressed. At the OCA level, the solutions however showed a slow convergence with a nonmonotonic evolution of the order parameter during the DMFT iterations. This could indicate that the true unconstrained anomalous distribution function of the steady-state is different from the imposed Fermi-function-like shape. Clarifying its precise form in the case of different effective doublon and holon temperatures is an interesting subject for future investigations. 

We also studied photo-doped nonequilibrium steady states in correlated metallic systems with $U<U_c$. Such states, with a half-filled quasi-particle band and some population of doublons and holons in the Hubbard satellites, could be potentially realized as short-lived transient states after photo-excitation. The question whether or not $\eta$-pairing exists in such systems is relevant in connection with the experiments reporting light-induced superconductivity in K$_3$C$_{60}$ and $\kappa$-organic compounds \cite{Mitrano2016,Buzzi2020,Budden2021}. Here, we found that nonequilibrium states with sufficiently high doublon and holon populations in the Hubbard bands and an inverted (negative temperature) distribution within the low-energy quasi-particle band can exhibit $\eta$-pairing at remarkably high temperature scales. On the other hand, we could not identify any $\eta$-pairing states in calculations which imposed a positive-temperature distribution within the quasi-particle band. 

A system with repulsive interactions and negative temperature distribution can be mapped onto one with attractive interactions and positive temperature distribution \cite{Tsuji2011}. Hence, if the population were completely inverted in a metallic state, we expect the emergence of uniform $s$-wave pairing (instead of staggered $\eta$-pairing), as indeed found in DMFT simulations \cite{Werner2019b}. The two Fermi-edge distribution for $U<U_c$ however corresponds to a nontrivial situation with a negative temperature for the quasi-particle states and positive effective temperatures for the photo-doped doublons and holons. Such a distribution can potentially be realized if the energy released by doublon-holon recombination processes is transferred to the quasi-particle states. It would be interesting to test this scenario in explicit time-dependent simulations of photo-doped metallic systems.  

\vspace{10mm}

\acknowledgements

The calculations were performed on the beo06 cluster at the University of Fribourg. PW and LG acknowledge support from the Swiss National Science Foundation via Grant No.~2000-1-240023. We thank M. Eckstein for helpful discussions.

\end{document}